\documentclass[preprint,12pt]{elsarticle}

\usepackage{graphicx}
\usepackage{amssymb}

\usepackage[hyphens]{url}
\usepackage{hyperref}
\usepackage{setspace}
\usepackage[linesnumbered,ruled,vlined]{algorithm2e}
\usepackage{algorithmic}
\usepackage{multirow}
\usepackage{geometry}
\usepackage{url}

\usepackage{subfigure}
\SetKwInput{KwInput}{Input}                
\SetKwInput{KwOutput}{Output}              

\journal{XXX}

\begin{document}
\setlength{\baselineskip}{20pt}

\begin{frontmatter}


\title{Conflict-Free Four-Dimensional Path Planning for Urban Air Mobility Considering Airspace Occupations}

\author[1stauthor,2ndauthor]{Wei Dai}
\author[1stauthor]{Bizhao Pang}
\author[1stauthor]{Kin Huat Low\corref{lkh}}
\cortext[lkh]{Corresponding author, email: mkhlow@ntu.edu.sg}

\address[1stauthor]{School of Mechanical and Aerospace Engineering, \\Nanyang Technological University, Singapore, 639798}
\address[2ndauthor]{Air Traffic Management Research Institute, \\Nanyang Technological University, Singapore, 637460}

\begin{abstract}
Urban air mobility (UAM) has attracted the attention of aircraft manufacturers, air navigation service providers and governments in recent years. Preventing the conflict among urban aircraft is crucial to UAM traffic safety, which is a key in enabling large scale UAM operation. Pre-flight conflict-free path planning can provide a strategic layer in the maintenance of safety performance, thus becomes an important element in UAM. This paper aims at tackling conflict-free path planning problem for UAM operation with a consideration of four-dimensional airspace management. In the first place, we introduced and extended a four-dimensional airspace management concept, AirMatrix. On the basis of AirMatrix, we formulated the shortest flight time path planning problem considering resolution of conflicts with both static and dynamic obstacles. A Conflict-Free A-Star algorithm was developed for planning four-dimensional paths based on first-come-first-served scheme. The algorithm contains a novel design of heuristic function as well as a conflict detection and resolution strategy. Numerical experiment was carried out in Jurong East area in Singapore, and the results show that the algorithm can generate paths resolving a significant number of potential conflicts in airspace utilization, with acceptable computational time and flight delay. The contributions of this study provide references for stakeholders to support the development of UAM.
\end{abstract}

\begin{keyword}
air traffic management \sep urban air mobility \sep strategic conflict-avoidance \sep path planning \sep A-Star algorithm


\end{keyword}

\end{frontmatter}


\section{Introduction}
\label{introduction}
    \subsection{Background and Motivation}
        Transportation, likened to be the blood of economic development, allows the resources to be provided to the places where they are needed. However, the land resource required in road transportation does not increase with the ballooning population and the traffic demand it brings in mega-cities. In recent years, as a result, the ascending frequency and severity of traffic congestion have become a major concern in the urban transport system. In the face of this problem, the utilization of the vertical dimension in the urban area is an intuitive choice for stakeholders. Starting from the second decade of the 21$^{st}$ century, attempts have been made to use low-altitude aircraft as a supplementary means of transport. Amazon and Google are the pioneers in the testing of drone deliveries, since which the interest in Unmanned Aerial Vehicle (UAV) operations has been growing \cite{Liu2021}. Benefit from the maturity of aircraft systems and related technologies, some companies started to look at not only cargo flights but also passenger air vehicles that are presented as electric vertical take-off and landing (eVTOL) aircraft. Airbus, Volocopters, EHang, Bell, and Embraer are developing their eVTOL aircraft, aiming at a leading position in this market \cite{Causa2021}. In this background, the concept of UAM was initiated, defined as ``safe and efficient air traffic operations in a metropolitan area for manned aircraft and unmanned aircraft systems (UAS)'' \cite{Thipphavong:2018}. 
    
        There will be a clear demand for UAM operations \cite{Ploetner2020}. Nevertheless, before large scale UAM operation can be realized, an emergent problem that has to be solved is ensuring the safety of UAM flight as they bring additional risk to the current air traffic system \cite{Wang2020}. The development of reliable urban aircraft is a challenge to the aviation industry \cite{Radmanesh2020}. Moreover, UAM as a new branch of airspace users leads to significant difficulties to air traffic management (ATM). Despite the existence of ATM system that has been operating safely and efficiently for many years, there are unignorable technical and management gaps that need to be coped with before UAM can be sufficiently integrated into the current air traffic system because the characteristics of UAM are different from those of traditional air traffic, which brings on obstacles in ensuring its operational safety and efficiency.

        To address these problems, AirMatrix was designed in 2017 as a comprehensive routing network for UTM \cite{Salleh2017}, and has been extended as a concept specifically for urban airspace management. In AirMatrix concept, urban airspace is discretized into blocks. By preventing conflicts in block utilization, the concept provides a solution for dynamic airspace utilization within a manageable overall airspace framework. 
        
        Path planning is one of the key problems in AirMatrix. In traditional manned aviation, pre-flight planning provides a path for the aircraft to follow. The planned path also serves as an estimation that determines the resources allocated to the flight, including fuel, flight crew, airspace resource, etc. In this case, the performance and reliability of pre-flight planning have an impact on the profitability of flight operators. In the context of UAM, high-density flight operation can be expected, thus it is unlikely that air navigation service providers (ANSP) will offer tactical air traffic control service to every aircraft in urban airspace. {There have been many studies on the development of on-board detect-and-avoid (DAA) systems, towards an autonomous collision avoidance capability that meets the target level of safety (TLS) of manned aviation, e.g. $10^{-7}$ fatalities per flight hour \cite{jamoom2016unmanned, shizhuang2021highly}. However, due to the complex environment that UAM flight operations in, many sources of interruptions  will cause uncertainties to the DAA system, which may impede the safety performance. As a result, there is no industrial practice showing that the TLS is achieved by DAA system independently. Therefore, multi-layered safety assurance is still crucial to UAM operation, especially in the early stage. }As a consequence, it is important that pre-flight path planning not only concerns high operational efficiency, but also provides a strategic layer of conflict avoidance. This requirement increases the difficulty of UAM path planning. 
    
    \subsection{Related Works}
        Path planning in urban airspace especially for drone operations has been studied based on various operational scenarios, such as delivery \cite{Liu2019}, surveillance \cite{Wu2021}, building inspection \cite{RAKHA2018252}. Many typical algorithms for path planning have been applied to aircraft applications, including sampling based algorithms like rapid exploring random tree (RRT) \cite{lavalle1998rapidly} and probabilistic road maps (PRM) \cite{kavraki1996probabilistic}, graph-based optimal algorithms like Dijkstra's algorithm \cite{dijkstra1959note} and A-Star (A*) \cite{hart1968formal, Li2020} algorithm, potential field method \cite{Shin2021}, optimal control \cite{anderson2010optimal}, population-based algorithms \cite{pehlivanoglu2012new, duan2010three}, etc. 
    
        In the context of path planning with a consideration of conflict avoidance, bio-inspired algorithms are widely used as this kind of algorithm is suitable for generating multiple paths at the same time and can easily implement complicated objectives and constraints \cite{8943975}. Existing methods include Particle Swarm Optimization (PSO) \cite{LIU20191504}, Ant Colony Optimization (ACO) \cite{duan2010three}, differential evolution (DE) \cite{nikolos2005coordinated}, Genetic Algorithms (GA) \cite{pehlivanoglu2012new}, Firefly Algorithm (FA) \cite{yang2010firefly}, memetic computing method \cite{iacca2013memory}, and other modified or improved algorithms based on them. However, these population-based methods have some weaknesses making them hard to be applied in UAM. For example, these methods are indeterministic and difficult to be analyzed, and their convergence rate is hard to guarantee \cite{WU2021100844}. 

        Graph-based algorithms are naturally suitable for path planning with a consideration of airspace management in which the airspace discretization has been performed and can be directly modeled as a node-and-link graph. A* algorithm is one of the most widely used graph-based algorithm, and has been applied in recent studies for aircraft path planning. Liu et al. \cite{liu2018search} proposed an A* planner for drone operating in a cluttered environment focusing on precise avoidance of collision with static obstacles considering attitude constraints. Penin et al. \cite{8543823} integrated perception constraints in A* algorithm for minimum-time trajectory planning. In their method, the uncertainties in state measurements are considered and updated in order to enhance the capability of collision avoidance. {These studies focuses on kinematic level motion planning for an individual aircraft in a cluttered area. The avoidance of vehicle-to-obstacle collision is achieved, but the vehicle-to-vehicle collision avoidance on the traffic management level is not considered in these studies.} Ma et al. \cite{7930420} presented a decentralized method for aircraft motion planning in high-density operations. Their algorithm combines an A* path planner and a low-level coordination strategy to achieve conflict avoidance among the drones. {Their design of local path planer appears to be similar to the DAA system, which makes the algorithm a combination of strategical and tactical planner. The algorithm was tested indoors, and the performance of the application of the algorithm in high-density drone operation in large urban area is unknown.} Tan et al. \cite{Tan2019} developed a framework combining an A* path planner and a scheduling scheme based on GA for conflict resolution. {This method resolves conflicts only by modifying departure times. Though not being considered in this paper, for a fully utilization of the four-dimensions of trajectory, detouring and hovering while the aircraft is airborne can be used in avoiding conflicts.}
    
    \subsection{Contributions of this paper}
    
        In this paper, conflict-free path planning problem is solved in a context of urban airspace. The main contributions of this paper is concluded as follows:
    
    \begin{itemize}
        \item A path planning model for conflict-free UAM operation is established. The model includes a point mass model for aircraft speed estimation in the shortest flight time path planning. Higher-priority flight plans are considered as dynamic obstacles, and mathematical representations for the avoidance of both static and dynamic obstacles are presented.
        \item A conflict-free A-star (CFA*) algorithm is developed for 4D path planning based on first-come-first-served (FCFS) scheme. The algorithm includes a novel design of heuristic function as well as a decision-making process for conflict avoidance.
        \item The algorithm was analyzed by a numerical experiment carried out in an urban airspace in Singapore. Results show that the algorithm {successfully} resolves a large number of potential conflicts with acceptable flight-time cost and computational time. 
    \end{itemize}

\subsection{{Organization of the paper}}

    {The rest of the paper is organized as follows. Section 2 builds the mathematical model of airspace under AirMatrix concept, formulates the path planning under AirMatrix concept as an optimization problem, and develops the model of aircraft velocity performances. Section 3 introduces the development of our new algorithm that solves the path planning problem. In Section 4, simulations using real-world building data are presented, and discussions are made based on the results. Section 5 concludes the findings in this study.}
    
\section{Problem Formulation}
    \subsection{AirMatrix Concept for Urban Airspace Management}
\begin{figure}[hbtp]
            \centering\includegraphics[width=\linewidth]{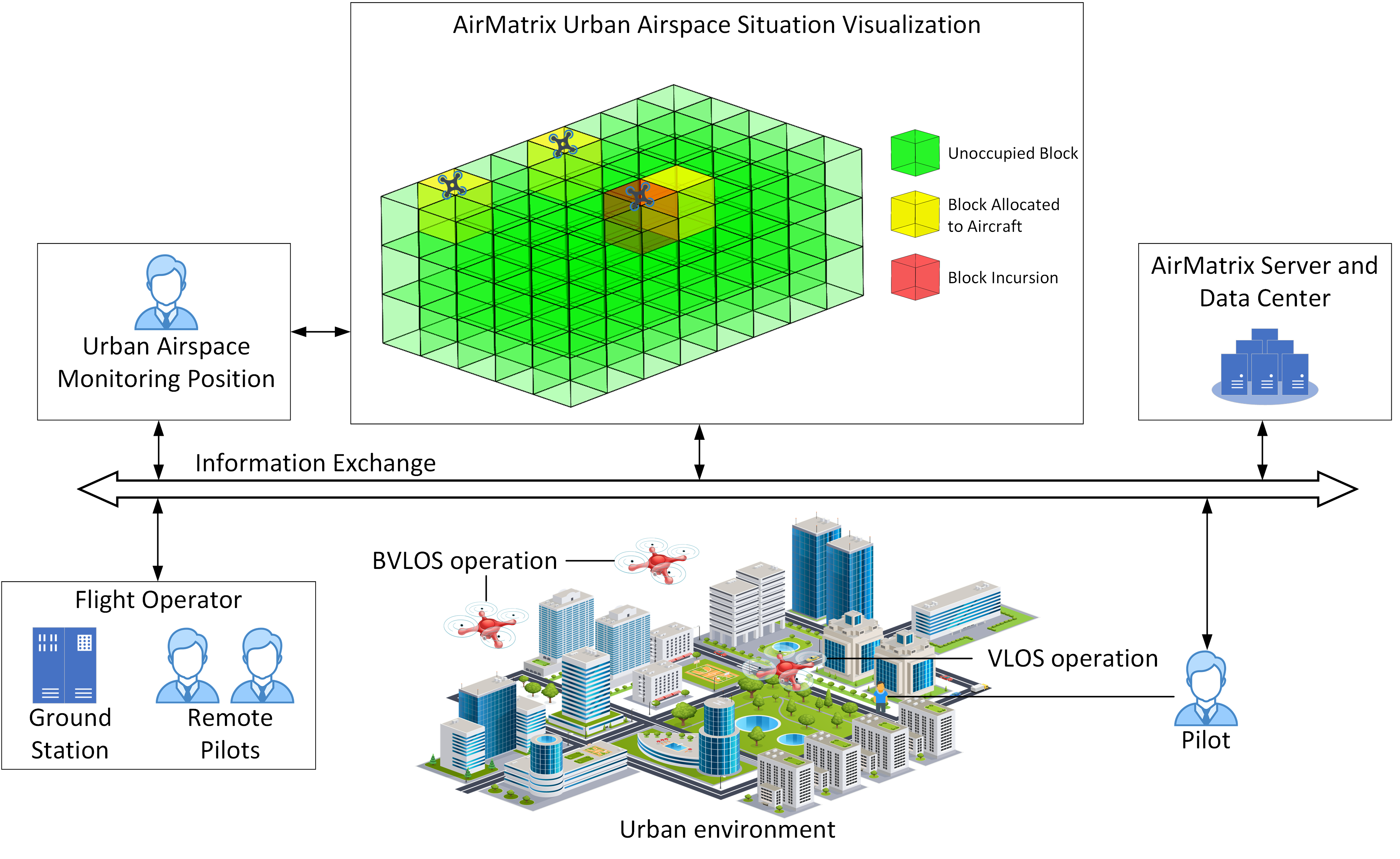}
            \caption{Four-dimensional airspace management based on AirMatrix concept}
            \label{airmatrix_overall}
        \end{figure}

        The AirMatrix serves for 4D airspace management, where a centralized service system is required to precisely record and manage the utilization of the airspace, as shown in Figure~\ref{airmatrix_overall}. The idea of AirMatrix concept is decomposing urban airspace into blocks of a certain size. There is a similar airspace management concept in the literature \cite{Pongsakornsathien}, which offers a higher resolution of airspace segmentation leading to higher computational consumption. In AirMatrix, flight operators are required to stream aircraft tracking information to the server which monitors the operational state of the entire airspace. The server recognizes normal and abnormal status of each block by comparing actual block utilization to the planned. {These status of blocks can be visualized as green, yellow and red blocks as shown in Figure~\ref{airmatrix_overall}, to provide situational awareness to UAM navigation service suppliers, who are on behalf of airspace surveillance and contingency management.}
        
        Thus the AirMatrix concept allows us to model the airspace as a three dimensional array of blocks, denoted as:
            \begin{equation}
                AirMatrix = \{B_{i,j,k}, i \in \{ 1,2, \dots, I\}, j \in \{ 1,2, \dots, J\}, k \in \{ 1,2, \dots, K\} \}
            \end{equation}
        where $B_{i,j,k}$ refers to a block indexed by $i$, $j$ and $k$, {meaning that the block is at the ith position in x-direction, the jth position in y-direction and kth position in z-direction in the entire AirMatrix network}. $I$, $J$, $K$ are the maximum values of $i$, $j$ and $k$, respectively, and are computed by the size of the modeled airspace and the size of each block. 
        
        The status of block occupation in AirMatrix can be presented as a four-dimensional array $Occup \in N^{I\times J\times K\times T}$, where each element $Occup_{i,j,k,t} \in \{0, 1\}$ denotes whether $B_{i,j,k}$ is occupied at time $t$, one means occupied and zero means not occupied. 
        
        Blocks that have overlaps with static obstacles like buildings should be recognized as occupied. In other words, these buildings permanently utilizes the blocks:
        \begin{equation}
            Occup(Building)_{i,j,k,t} = \left\{
            \begin{array}{lr}
            1,     & \bigcup_{Building} B_{i,j,k}\cap Building \ne \o, t \in T\\
            0,     & \bigcup_{Building} B_{i,j,k}\cap Building = \o, t \in T
            \end{array}
            \right.
        \end{equation}

    \subsection{Objective of Optimization}
    
        Under AirMatrix concept, the objective of path planning problem is formulated as:
        
        \begin{equation}
	        \mathop{\arg\min}_{Path} \sum_{B^{(a)} \in Path}Cost(B^{(a)})
        \end{equation}
        where $Path = [\begin{array}{cccccc}
            B^{(1)}= B^{start}     &  B^{(2)} & B^{(3)} & \dots & B^{(n-1)} & B^{(n)}= B^{end}
            \end{array}]$ is the path from the starting point to the end point, represented by a list of blocks. {The superscript of $B$ refers to the ordinal number of this block in a planned path. $B^{start}$ refers to the block where the start point is, $B_{end}$ refers to the block where the end point is, and $B^{(a)}$ refers to the a$^{th}$ block in a planned path, where $a=1,2,3,…,n$. And we have $B^{(1)} = B^{start}$, $B^{(n)}=B^{end}$ according to the definition. $B^{start}$ and $B^{end}$ are denoted as:}
            
        \begin{equation}
            B^{start} = \mathop{\arg\min}_{B} \left|(x,y,z)^{B},(x, y, z)^{(take-off)} \right|, \quad B \in AirMatrix
        \end{equation}
        \begin{equation}
            B^{end} = \mathop{\arg\min}_{B} \left|(x,y,z)^{B},(x, y, z)^{(landing)} \right|, \quad B \in AirMatrix
        \end{equation}
        where operator $|\ ,\ |$ is defined to compute the distance between two locations. $x$, $y$ and $z$ are the coordinates in a Cartesian coordinate system. They refer to the location of center point when indexed by a block. 
        
        In this research, the objective is to minimized the total flight time. Therefore the cost at each block is denoted as: 

        \begin{equation}
	        Cost(B^{(a)})
	        =\left\{
             \begin{array}{ll}
             \frac{1}{2}\cdot \frac{|B^{(a)},B^{(a+1)}|}{v^{a,a+1}} + t^{(a)}_{hover}, & a=1 \\
             \frac{1}{2}\cdot(\frac{|B^{(a-1)},B^{(a)}|}{v^{a-1,a}} + \frac{|B^{(a)},B^{(a+1)}|}{v^{a,a+1}})+t^{(a)}_{hover}, & 1<a<n \\
             \frac{1}{2}\cdot \frac{|B^{(a-1)},B^{(a)}|}{v^{a-1,a}} + t^{(a)}_{hover}, & a=n
             \end{array}
             \right.
        \end{equation}
        where $v^{a,b}$ denotes the aircraft velocity between $B^{(a)}$ and $B^{(b)}$. 
    
    \subsection{Aircraft Velocity Estimation}
    \label{velocity}
        In the AirMatrix concept, aircraft movement is not limited to vertical or horizontal. Each block is connected to 26 neighbours by links with different elevation angles. However, most multirotor manufacturers do not publish the velocity of climbing with a path angle $\varphi$ in their aircraft specifications. In order to estimate these velocities, we used the specifications given by aircraft manufacturers, including aircraft mass, max vertical speed, and max horizontal speed, to build a simplified mass-point model, given in:
        \begin{equation}
            m \dot{\textbf{v}}= \textbf{d} + m [\begin{array}{ccc} 0 & 0 & -g \end{array}]^T + \textbf{T}_h \label{motion}
        \end{equation}
        where $m$ is mass, $v$ is velocity denoted as $\textbf{v}=[\begin{array}{ccc} \dot{v}_x & \dot{v}_y & \dot{v}_z \end{array}]^T$, $\textbf{T}_h$ is thrust denoted as $\textbf{T}_h = [\begin{array}{ccc} T_{hx} & T_{hy} & T_{hz} \end{array}]^T$. $\textbf{d}$ is drag defined as $\textbf{d} = [\begin{array}{ccc} {d}_x & {d}_y & {d}_z \end{array}]^T$. {In fluid dynamics, the magnitude of drag is formulated as $d = 1/2\rho v^2C_DA$, where $\rho$ is the density of the fluid, $C_D$ is the drag coefficient, and $A$ is the reference area. However, $\rho$ and $A$ change as the path angle changes, making it difficult to estimate the speed performance. In this study, we are not aiming at building a perfectly accurate aircraft performance model, but determining a flyable speed that could be allocated to the flight from the perspective of traffic management. To simplify the computation without losing generality, we assume that $d = ev^2$, where $e = 1/2\rho C_DA$ is a constant subject to an aircraft type.}
        
        \begin{figure}[hbtp]
            \centering\includegraphics[width=0.5\linewidth]{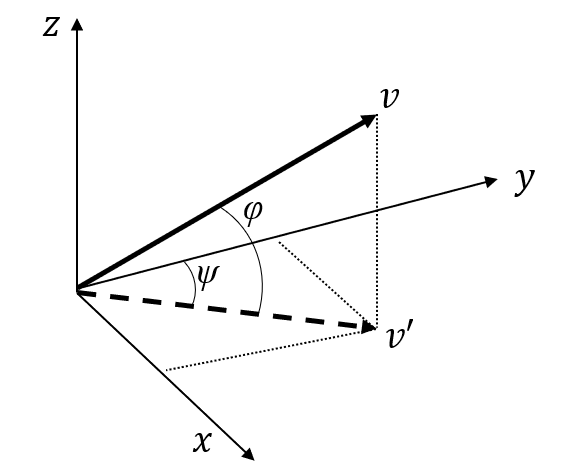}
            \caption{Velocity decomposition}
            \label{v_decomp}
        \end{figure}
        
        Decomposition of velocity is shown in Figure~\ref{v_decomp}, where $\textbf{v'}$ is the projection of $\textbf{v}$ in horizontal plane, and $\varphi$ is the angle between $\textbf{v}$ and $\textbf{v'}$. $\psi$ is the angle between $\textbf{v'}$ and y-axis. In the condition of maximum speed vertical climbing, $\varphi=\pi/2$. In this case, the speed $\textbf{v}_{mv}$ and drag $\textbf{d}_{mv}$ are:
        
        \begin{equation}
            \textbf{v}_{mv} = [\begin{array}{ccc} 0 & 0 & max(v_{vertical}) \end{array}]^T
        \end{equation}
        
        \begin{equation}
            \textbf{d}_{mv} = [\begin{array}{ccc} 0 & 0 & -e \cdot max(v_{vertical})^2 \end{array}]^T
        \end{equation}
        In the condition of maximum speed horizontal cruising, $\varphi=0$, hence the speed $\textbf{v}_{mh}$ and drag $\textbf{d}_{mh}$ in this case are:
        \begin{equation}
            \textbf{v}_{mh} = [\begin{array}{ccc} sin(\psi) \cdot max(v_{horizontal}) & cos(\psi) \cdot max(v_{horizontal}) & 0 \end{array}]^T
        \end{equation}
        \begin{equation}
            \textbf{d}_{mh} = [\begin{array}{ccc} -e \cdot sin(\psi) \cdot max(v_{horizontal})^2 & -e \cdot cos(\psi) \cdot max(v_{horizontal})^2 & 0 \end{array}]^T
        \end{equation}
        The maximum power output of propeller system in both cases can be estimated as the dot product of thrust and velocity:
        \begin{equation}
            P_{max} = \textbf{T}_h^T \textbf{v}\label{max_power}
        \end{equation}
        In both cases, the aircraft in the state of uniform motion in a line, i.e. $\dot \textbf{v} = 0$. Combining Equation~(\ref{motion}) to~(\ref{max_power}), $P_{max}$ and $e$ can be computed, and velocity with any given $\varphi$ and $\psi$ value can be estimated. 
        
        {To verify the model, we take DJI Inspire 2 as a reference, which is the only aircraft type of which the tilt velocity performance is published by the manufacturer \cite{inspire}. The mass of DJI Inspire 2 is 3.44 kg, the maximum ascent speed at S-mode is 6 m/s, and the maximum cruise speed is 26 m/s. Therefore $e=0.0115$ and $P_{max}=202.479$ can be computed based on our model. The maximum tilt speed can be computed with a given path angle $\varphi$. When $\varphi$ is between 0 and 90$^\circ$, the maximum speed is between 6 and 26 m/s. The reference tilt speed given by the manufacturer is 4-9 m/s. This item is given in the section of ascent and descent speed, so that it is likely that the speed is referring to the situation that the path angle is large, i.e. $\varphi > 45^\circ$. The computed maximum speed corresponding to $\varphi = 45^\circ$ and 60$^\circ$ are 7.973 m/s and 6.682 m/s, respectively. Both of them fall in the interval given by the manufacturer, which confirms the effectiveness of our model. }    
    \subsection{Conflict Avoidance}
        Conflict avoidance in AirMatrix is based on preventing duplicate block utilization. For a planned $Path = [\begin{array}{ccccc}
        B^{(1)} & B^{(2)} & B^{(3)} & \dots & B^{(n)}\end{array}]$, the duration of occupation time in a block equals to the cost in the same block. And the time that the aircraft enters and leaves a block can be defined as:
        \begin{equation}
            t_{enter}(B^{(a)}) = 
            \left\{ \begin{array}{ll}
            t_{take-off}, & a=1\\
            t_{take-off} + \sum_{l=1}^{(a-1)} Cost(B^{(l)}), & a>1
            \end{array}
            \right.
        \end{equation}
        \begin{equation}
            t_{exit}(B^{(a)}) = t_{enter}(B^{(a)}) + Cost(B^{(a)})
        \end{equation}
        Then the occupation of each block is presented as $Occup(B^{(a)}) \in N^{I\times J\times K\times T}$:
        \begin{equation}
            Occup(B^{(a)})_{i_a, j_a, k_a, t} = 1, \quad t \in [t_{enter}(B^{(a)}), t_{exit}(B^{(a)})]
        \end{equation}
        where $i_a$, $j_a$, and $k_a$ are the indexes of $Block^{(a)}$. The block occupation of a planned path is:
        \begin{equation}
            Occup(Path) = \bigcup_{B \in Path} Occup(B)
        \end{equation}
        The summation of all block utilization by planned paths and buildings is:
        \begin{equation}
            Occup(total) = \sum_{Path} Occup(Path) + Occup(Building),\ Path\in \{all\ planned\ paths\}
        \end{equation}
        The condition of no conflict can be formulated as:
        \begin{equation}
        \begin{array}{c}
        \forall i \in \{ 1,2, \dots, I\}, \forall j \in \{ 1,2, \dots, J\}, \forall k \in \{ 1,2, \dots, K\},\forall t \in \{ 1,2, \dots, T\}\\
            Occup(total)_{i,j,k,t} \le 1,\ 
            \end{array}\label{conflict}
        \end{equation}
        
\section{Conflict-Free A-Star Algorithm for Path Planning}
    \subsection{Heuristic Function Design}
    A* algorithm can be seen as an extension of Dijkstra algorithm \cite{dijkstra1959note}, for its introduction of a heuristic searching on the basis of Dijkstra algorithm. In A* algorithm, a heuristic function presents an estimated cost from a node to the destination. Every node $n_i$ has two attributes: cost function $G(n_i)$ denoting the cost from starting node, and heuristic function $H(n_i)$ denoting the estimated cost from node $n_i$ to the destination node. At each iteration step, the algorithm searches the neighbours of the node with the least $F(n_i)$ value, where $F(n_i) = G(n_i) + H(n_i)$. {The nodes close to the destination have smaller $H$ values. Compared with Dijkstra algorithm which searches the nodes with smallest $G$ value, A* algorithm searches the nodes with the smallest $F$ value at each step, thus the nodes closer to the destination will be searched earlier. This feature reduces the required number of searching steps in A* algorithm compared with Dijkstra algorithm, making the A* algorithm faster.}
    
    The heuristic function $H(n_i)$ is normally defined as the Euclidean distance or Manhattan distance between $n_i$ and the destination node $n_D$. {In shortest-flight-time trajectory planning, the cost to be minimized in the algorithm is flight-time. Due to the performance of the aircraft, vertical velocity and horizontal velocity are significantly different. Therefore the same distance in vertical and horizontal direction will lead to a very different time cost. Thus Manhattan or Euclidean distance are no longer a proper estimation of the cost from a node to the destination.} Therefore a modified heuristic function based on the optimization objective is needed. The design of heuristic function affects the performance of A* algorithm. If the heuristic value is larger than actual cost, the algorithm will search faster but the optimality is not guaranteed. On the other hand, if the heuristic value is smaller than the actual cost, the optimality can be achieved while the searching time of the algorithm will be prolonged. The best heuristic function should be close to but not larger than the actual cost. In this study, a novel heuristic function was designed for shortest-flight-time trajectory searching in a 3D environment based on the speed estimation model introduced in Section~\ref{velocity}.
    
    
    In the path planning in AirMatrix network, a node represents the center of a block, aircraft is allowed to fly along a link that connect a block with one of its neighbour. The heuristic function shows an estimation of the cost from node $n_i$ to the destination node, hence it can be defined as the summation of the flight time spent on the links from $n_i$ to the destination node $n_D$:
    \begin{equation}
        H^*(n_i) = \sum_{j \in C}\frac{l_j}{v_j}
    \end{equation}
    where $l_j$ is the length of link j, $v_j$ is the velocity along link j, and C is the set of the required links from $n_i$ to $n_D$ in the shortest path, defined as the shortest route in AirMatrix network disregarding static and dynamic obstacles. The number of links' types, in terms of length and elevation angle, is finite. Thus the heuristic function can be formulated as:
    \begin{equation}
        H^*(n_i) = \sum_{k}c_k t_k \label{heuristic_function}
    \end{equation}
    where $t_k$ is the time that the aircraft takes to fly through a link referring to type k, and $t_k = l_k/v_k$. And $c_k$ is the required number of links referring to type k in the shortest path from $n_i$ to $n_D$. 
       
\begin{figure}[]
            \centering\includegraphics[width=0.4\linewidth]{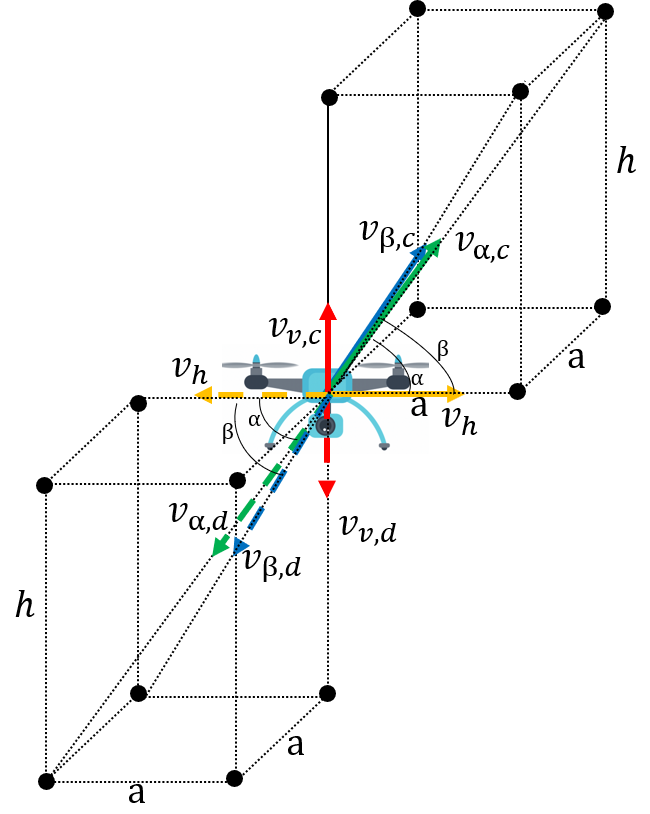}
            \caption{Velocities referring to selectable elevation angles of an aircraft at each time step}
            \label{typical_speeds}
       \end{figure}

    In AirMatrix, the length and width of a block, denoting dimensions in two horizontal directions, have the same impact on the efficiency and safety performance of UAM traffic. It's reasonable to set the length equals to the width, i.e. the bottom surface of each block can be a square. Because of the symmetry of AirMatrix network, the 26 links for a given node fall into 7 types with different elevation angles ($0^{\circ}$, $\pm \alpha$, $\pm \beta$, and $\pm 90^{\circ}$), as shown in Figure~\ref{typical_speeds}, where a refers to the side length of the bottom of AirMatrix block, h refers to the height of AirMatrix block, footnote $v$ refers to vertical, $h$ refers horizontal, $\alpha$ and $\beta$ refer to two elevation angles, $c$ and $d$ refer to climb and descend, respectively. $\alpha = sin^{-1}\sqrt{a^2+h^2}$ and $\beta = sin^{-1}\sqrt{2a^2+h^2}$. The speed performance with a negative elevation angle is difficult to estimate. {Because when the aircraft is tilt descending, it is not flying with max power. Thus, the max velocity depends on the setting of the flight controller, not physical models. The controller setting varies significantly because building a drone is too easy nowadays.} In this study, we assume that $v_{\alpha,d}=v_{\alpha,c}$, $v_{\beta,d}=v_{\beta,c}$, and $v_{v,d}=v_{v,c}$, then the number of different types of speed reduces to four, denoted as $v_h$, $v_v$, $v_\alpha$, and $v_\beta$. These four speeds can be computed when the length, width, and height of each AirMatrix block is defined. The length of each link can also be computed. Thus $c_k$ is the only unknown term in Equation~(\ref{heuristic_function}).
       
\begin{figure}[]
            \centering\includegraphics[width=0.7\linewidth]{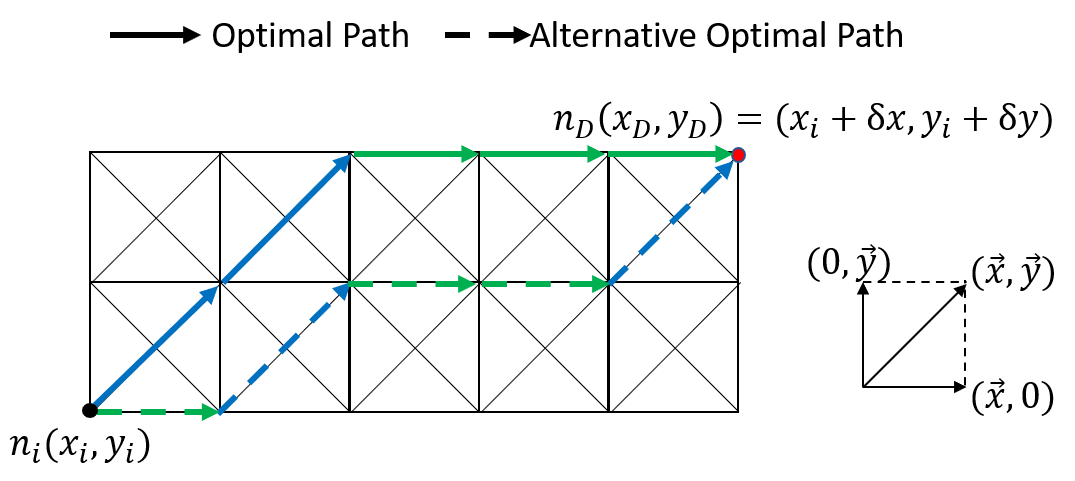}
            \caption{Optimal path in 2D scenario}
            \label{heuristic_2d}
        \end{figure}

    We firstly explain the computation of $c_k$ in a two-dimensional case, as shown in Figure~\ref{heuristic_2d}, where the black node $n_i$ is indexed by $(x_i,y_i)$, indicating the number of blocks from the origin node. The red node $n_D$ is the destination, indexed by $(x_D, y_D)$. $\delta x$ and $\delta y$ is the difference between $n_i$ and $n_D$, and in this figure, $\delta x = 5$, $\delta y = 2$. The solid arrows show an optimal path from $n_i$ to $n_D$ in the network. It consists of arrows in two different directions, illustrated in blue and green. The direction of a blue arrow is marked as $(\vec{x},\vec{y})$ and the direction of a green arrow is marked as $(\vec{x},0)$. There are alternative optimal paths, for example, the path shown by the dashed arrows. Every optimal path has the same number of arrows in each direction. From $n_i$ to $n_D$, a total movement of $(\delta x,\delta y)$ is required. Since one movement in $(\vec{x},\vec{y})$ direction is shorter than the summation of one in $(\vec{x},0)$ and one in $(0,\vec{y})$. An optimal path requires a maximum number of moves in $(\vec{x},\vec{y})$ direction, which is $min(\delta x,\delta y) = \delta y$. After that, the required movement is $(\delta x,\delta y)-\delta y(\vec{x},\vec{y})=(\delta x-\delta y,0)$, which is $\delta x-\delta y$ moves in $(\vec{x},0)$ direction. The heuristic value, presenting the flight time from $n_i$ to $n_D$ is:
       \begin{equation}
           H^*(n_i) = \delta y \cdot t_{\vec{x},\vec{y}}+(\delta x-\delta y)t_{\vec{x},0}
       \end{equation}
       where $t_{\vec{x},\vec{y}}$ and $t_{\vec{x},0}$ refer to the flight time of one movement in direction $(\vec{x},\vec{y})$ and $(\vec{x},0)$, respectively. To generalize this process, the computation is shown in Algorithm~\ref{heuristic_2d_algorithm}. 
       
       \begin{algorithm}
            \label{heuristic_2d_algorithm}
            \DontPrintSemicolon
            
            \If{$\delta x \geq \delta y$}{
                $H^*(n_i) = \delta y \cdot  t_{\vec{x},\vec{y}}+(\delta x-\delta y) t_{\vec{x},0}$
            }
            
            \If{$\delta x \textless \delta y$}{
                $H^*(n_i) = \delta x \cdot  t_{\vec{x},\vec{y}}+(\delta y-\delta x) t_{0,\vec{y}}$
            }
            return ($H^*(n_i)$)
            \caption{Heuristic Function in Two-Dimensional Case}
        \end{algorithm}
       
       This computation can be extended to the three-dimensional case. There are totally 7 different directions in the 3D network: $(\vec{x},0,0)$, $(0,\vec{y},0)$, $(0,0,\vec{z})$, $(\vec{x},\vec{y},0)$, $(0,\vec{y},\vec{z})$, $(\vec{x},0,\vec{z})$, and $(\vec{x},\vec{y},\vec{z})$. The computation is presented in Algorithm~\ref{heuristic_3d_algorithm}.
       \begin{algorithm}
            \label{heuristic_3d_algorithm}
            \DontPrintSemicolon
    
            
            \If{$\delta x \geq \delta y \geq \delta z$}{
                $H^*(n_i) = \delta z \cdot  t_{\vec{x},\vec{y},\vec{z}}+(\delta y-\delta z) t_{\vec{x},\vec{y},0}+(\delta x-\delta y) t_{\vec{x},0,0}$
            }
            \If{$\delta x \geq \delta z \geq \delta y$}{
                $H^*(n_i) = \delta y \cdot  t_{\vec{x},\vec{y},\vec{z}}+(\delta z-\delta y) t_{\vec{x},0,\vec{z}}+(\delta x-\delta z) t_{\vec{x},0,0}$
            }
            \If{$\delta y \geq \delta x \geq \delta z$}{
                $H^*(n_i) = \delta z \cdot  t_{\vec{x},\vec{y},\vec{z}}+(\delta x-\delta z) t_{\vec{x},\vec{y},0}+(\delta y-\delta x) t_{0,\vec{y},0}$
            }
            \If{$\delta y \geq \delta z \geq \delta x$}{
                $H^*(n_i) = \delta x \cdot  t_{\vec{x},\vec{y},\vec{z}}+(\delta z-\delta x) t_{0,\vec{y},\vec{z}}+(\delta y-\delta z) t_{0,\vec{y},0}$
            }
            \If{$\delta z \geq \delta x \geq \delta y$}{
                $H^*(n_i) = \delta y \cdot  t_{\vec{x},\vec{y},\vec{z}}+(\delta x-\delta y) t_{\vec{x},0,\vec{z}}+(\delta z-\delta x) t_{0,0,\vec{z}}$
            }
            \If{$\delta z \geq \delta y \geq \delta x$}{
                $H^*(n_i) = \delta x \cdot  t_{\vec{x},\vec{y},\vec{z}}+(\delta y-\delta x) t_{0,\vec{y},\vec{z}}+(\delta z-\delta y) t_{0,0,\vec{z}}$
            }

            return ($H^*(n_i)$)
            \caption{Heuristic Function in Three-Dimensional Case}
        \end{algorithm}
        
        \subsection{Strategic conflict avoidance}
        Strategic conflict avoidance is realized by avoiding duplicate block utilization. When an occupied block is detected in a searching step of the algorithm, there are two options to avoid the conflict. Either hovering at the previous block or routing through another block can solve the conflict, and a decision-making process is required to decide which solution will be used. In this study, the process is achieved by a greedy algorithm. {The greedy algorithm is an algorithm that takes the optimal choice at each stage. Therefore, the algorithm doesn't produce a global optimal solution, but can yield locally optimal solutions that approximate a globally optimal solution in a reasonable amount of time \cite{cormen2009introduction}.} This CFA* algorithm leads to sub-optimality in the entire trajectory planning. The pseudocode of CFA* algorithm is presented in Algorithm~\ref{algorithm}. {In step 14, the enter and exit time of the node under searching is estimated by using the velocity computed with the model developed in Section 2.3. In step 15 of the algorithm, conflict detection is performed based on Equation~\ref{conflict}. If a conflict is detected, the algorithm computes the required duration of the hovering at the previous node to solve the conflict (step 16), and identifies if the hovering is available (step 18), i.e., won't cost a secondary conflict. If there is an available hovering to solve the conflict, the cost of hovering is added to the cost of the node (step 21). And the duration of hovering is recorded in step 23 and will be used in generating the entire path.}
        
        \begin{algorithm}
            \label{algorithm}
            \DontPrintSemicolon
                OpenList = \o\\
                CloseList = \o\\
                add starting node to OpenList\\
                \While{OpenList $\neq$ \o}
                {
                    minF $\leftarrow$ the node with least F value in OpenList\\
                    /*\textit{F = G + H, where G is the cost function, H is the heuristic function}*/\\
                    remove minF from OpenList\\
                    add minF to CloseList\\
                    find the neighbours of minF and set their parents to minF\\
                    \For{node in neighbours}
                    {
                        \If{node is in the CloseList}
                        {
                            continue\\
                        }
                        compute node.G\\
                        compute the time entering and exiting node\\
                        \If{conflict is detected}
                        {
                            compute the required HoveringTime for avoiding conflict\\
                            /* \textit{if hovering not available}*/\\
                            \If{HoveringTime larger than threshold \textbf{or} HoveringTime is not available due to conflict at minF}
                            {
                                continue\\ 
                            }
                            /* \textit{if hovering available}*/\\
                            node.G += HoveringTime\\
                            node.AfterHovering = True\\
                            node.HoveringTime = HoveringTime\\
                        }
                        \If{node == ExistNode in the OpenList}
                            {
                                
                                \If{node.G $<$ ExistNode.G}
                                {
                                    replace ExistNode by node\\
                                }
                            }
                        \Else{add node to OpenList}
                    }
                    \If{ending node in CloseList}
                    {
                        return True
                    }
                }
                return False
            \caption{Conflict-Free A-Star Algorithm in AirMatrix}
        \end{algorithm}

\section{Simulations and Results}
    \subsection{Block Sizing in AirMatrix}
    Block size in AirMatrix is a major factor influencing flight safety performance, as it reflects horizontal and vertical separation between two aircraft. The ability of an aircraft to maintain on the pre-defined path and minimize deviation is the main consideration in the determination of block sizing. Such deviation, named as total system error (TSE), consists of path definition error (PDE), flight technical error (FTE) and navigation system error (NSE) \cite{ICAO2008}. The PDE can be assumed to be negligible \cite{ICAO2008}, and according to our previous study, the lateral deviation due to FTE is minor \cite{Wang2020_2}. 
    
\begin{figure}[hbtp]
            \centering\includegraphics[width=0.6\linewidth]{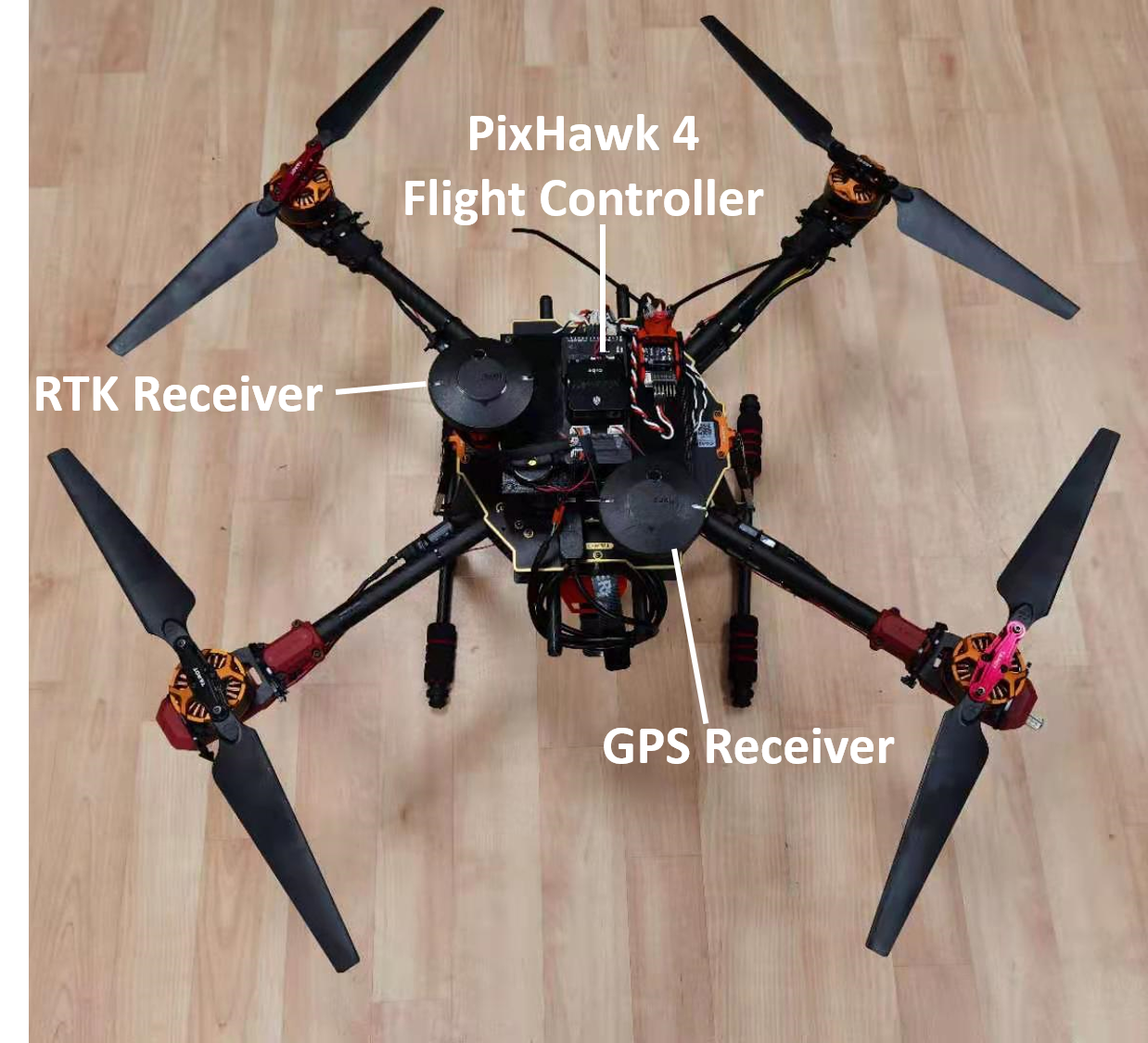}
            \caption{Drone and sensors set up for experiments}
            \label{drone}
    \end{figure}

    In order provide an estimation of NSE, an experiment was carried out with a self-built drone equipped with Real-Time Kinematics (RTK) as shown in Figure~\ref{drone}. RTK provides tracking information with $\pm 0.02$ meters accuracy. The aircraft is navigated by PixHawk 4 where multi-sensor data fusion is performed by an Extended Kalman Filter (EKF) to provide localization. In this case, RTK output is regarded as true value, and is compared the tracking record of PixHawk 4 to analyze NSE. The experiment contains 65 flights and the total flight time is 3.83 hours. 
    
\begin{figure*}[h]
          \centering
            \subfigure[Horizontal error]{\includegraphics[width=0.45\textwidth]{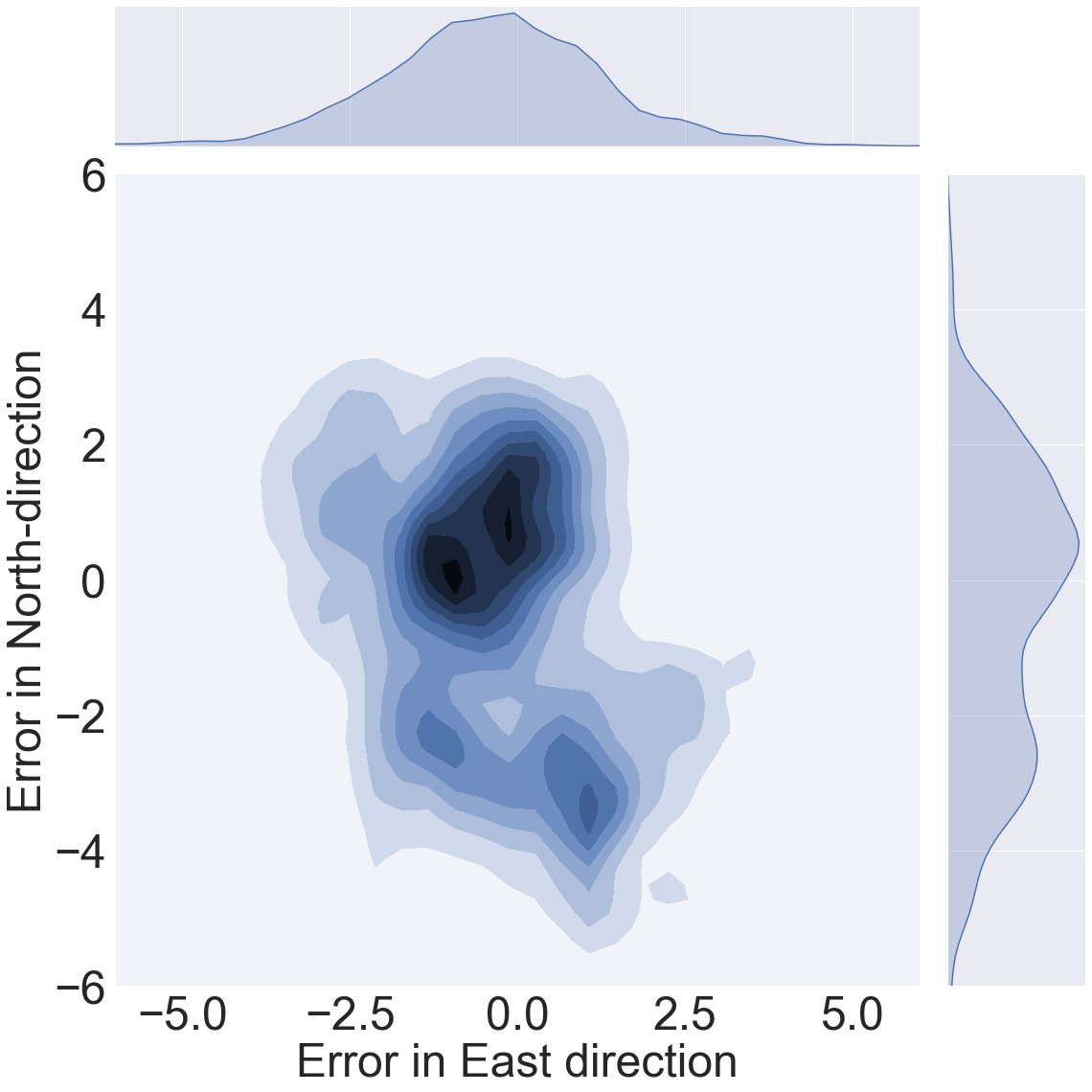}}
        	\subfigure[Vertical error]{\includegraphics[width=0.45\textwidth]{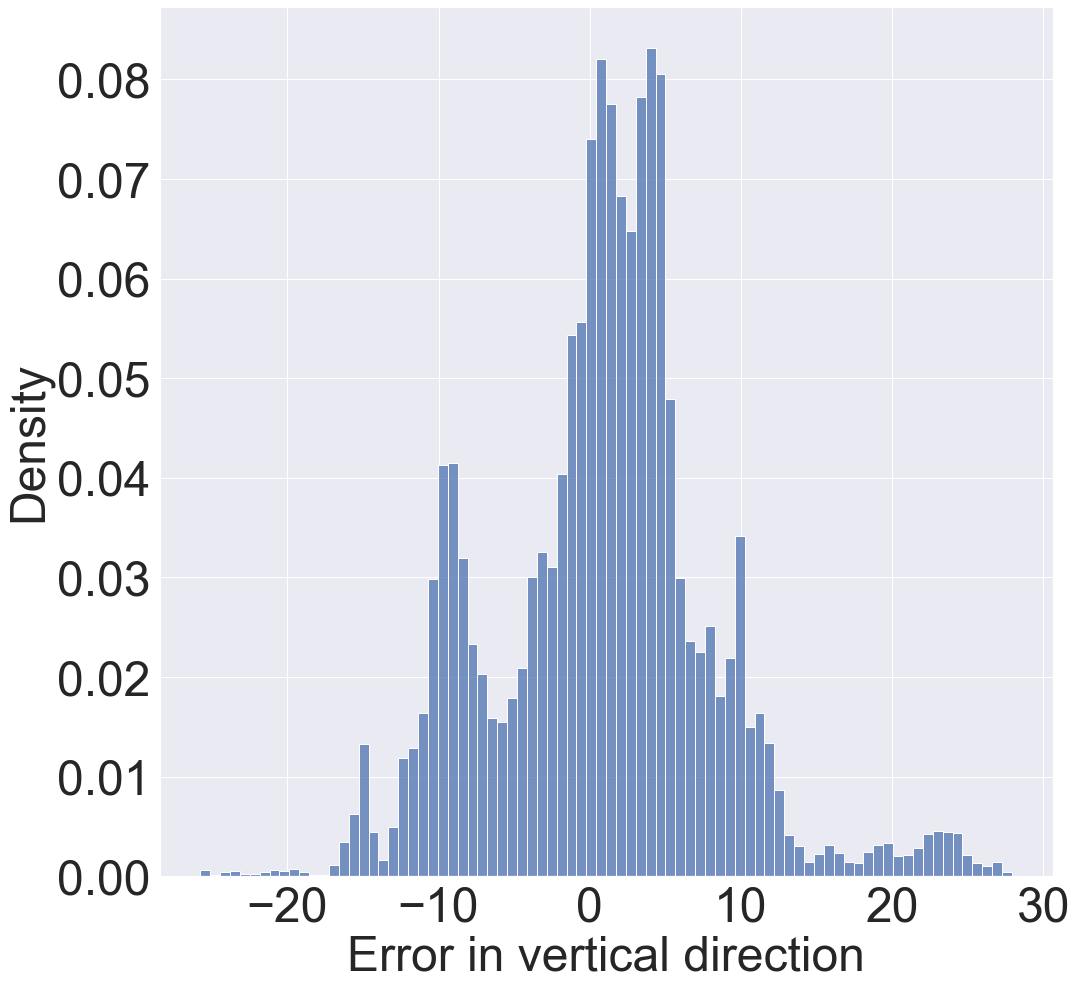}}
          \caption{Positioning error of EKF fused data}
        	\label{gps_error}
        	\vspace{0.3in}
        \end{figure*}

    
    Experiment results are illustrated in Figure~\ref{gps_error}(a) and (b). Overall the errors in horizontal and vertical directions all present Gaussian-like distributions. As further analysis of NSE error is not the focus of this paper, we can perform block sizing selection with the statistics of NSE given in Table~\ref{NSE}. In this paper, the length and width of block are 20 meters. Considering path planning are based on the centers of the blocks, this will lead to a confidence of higher than 99\% that the aircraft flies within the block allocated to it in the horizontal dimension. {It is worth noting that the current set of our drone is using a combination of barometer and IMU. Under this set, the on-board altitude measurement is influenced by the wind, and the drone deviates from its assigned height. Meanwhile, there are various sources of GNSS inference in urban area, and our experiment data shows that the performance of GNSS is statistically worse. Therefore, barometer and IMU were used for altitude measurement in this study. Nevertheless, the performance of GNSS is easier to be improved compared with that of barometer. There have been many studies on ground-based augmentation system (GBAS) for GNSS, that can be implemented for UAM operations. It can be foreseen that with the deployment of GBAS, GNSS will provide a higher accuracy for UAM in the future.} In this study, we assume that 99\% confidence can be expected if the block height is set to 40 meters. By setting the maximum altitude limit to 500 feet (150 meters), the AirMatrix built under the height limit has three horizontal layers.
    
    \begin{table}[!h]
    \begin{center}
    \caption{Statistics of NSE}
        \label{NSE}
        \begin{tabular}{lcccc}
        \hline
        \multicolumn{1}{c}{\textbf{}} & \multirow{2}{*}{\textbf{Mean (m)}} & \multirow{2}{*}{\textbf{\begin{tabular}[c]{@{}c@{}}Standard\\ Deviation (m)\end{tabular}}} & \multicolumn{2}{c}{\textbf{Error (m)}} \\
        \multicolumn{1}{c}{}          &                   &                   & \textbf{0.5 Percentile}       & \textbf{99.5 Percentile}       \\ \hline
        Error in East       & -0.289             & 2.137              & -6.137             & 7.844             \\
        Error in North      & -0.578             & 2.299              & -7.843             & 4.885             \\
        Horizontal error              & 2.650              & 1.802              & 0.079              & 10.130            \\
        Vertical error                & 0.831              & 7.353              & -16.206            & 24.432            \\ \hline
        \end{tabular}
    \end{center}
    \end{table}
    
    \subsection{Simulation Scenario}
\begin{figure*}[]
          \centering
            \subfigure[An urban area in Jurong East, Singapore]{\includegraphics[width=0.6\textwidth]{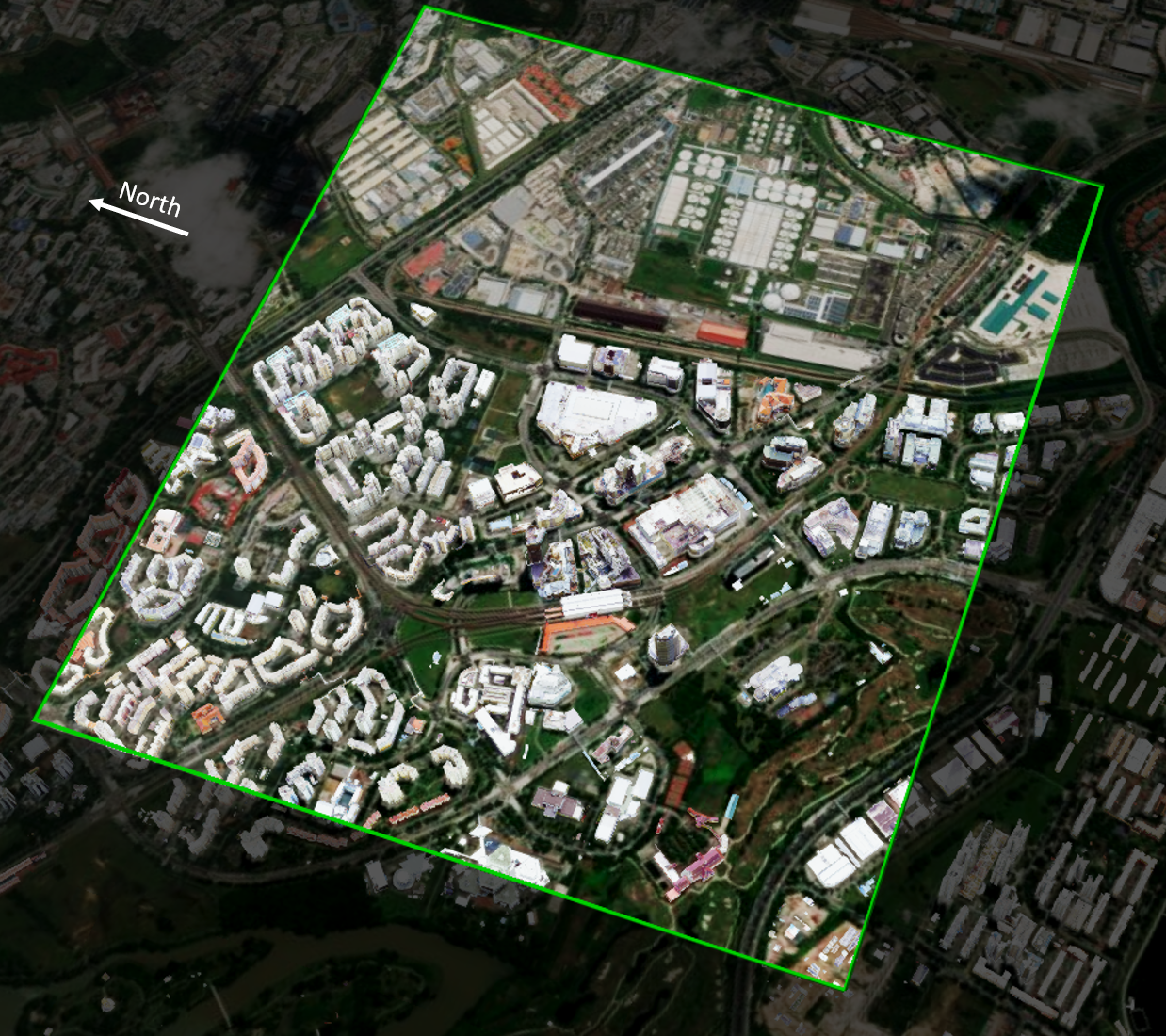}}\label{jurong_east} \\
        	\subfigure[Blocks occupied by buildings]{\includegraphics[width=0.6\textwidth]{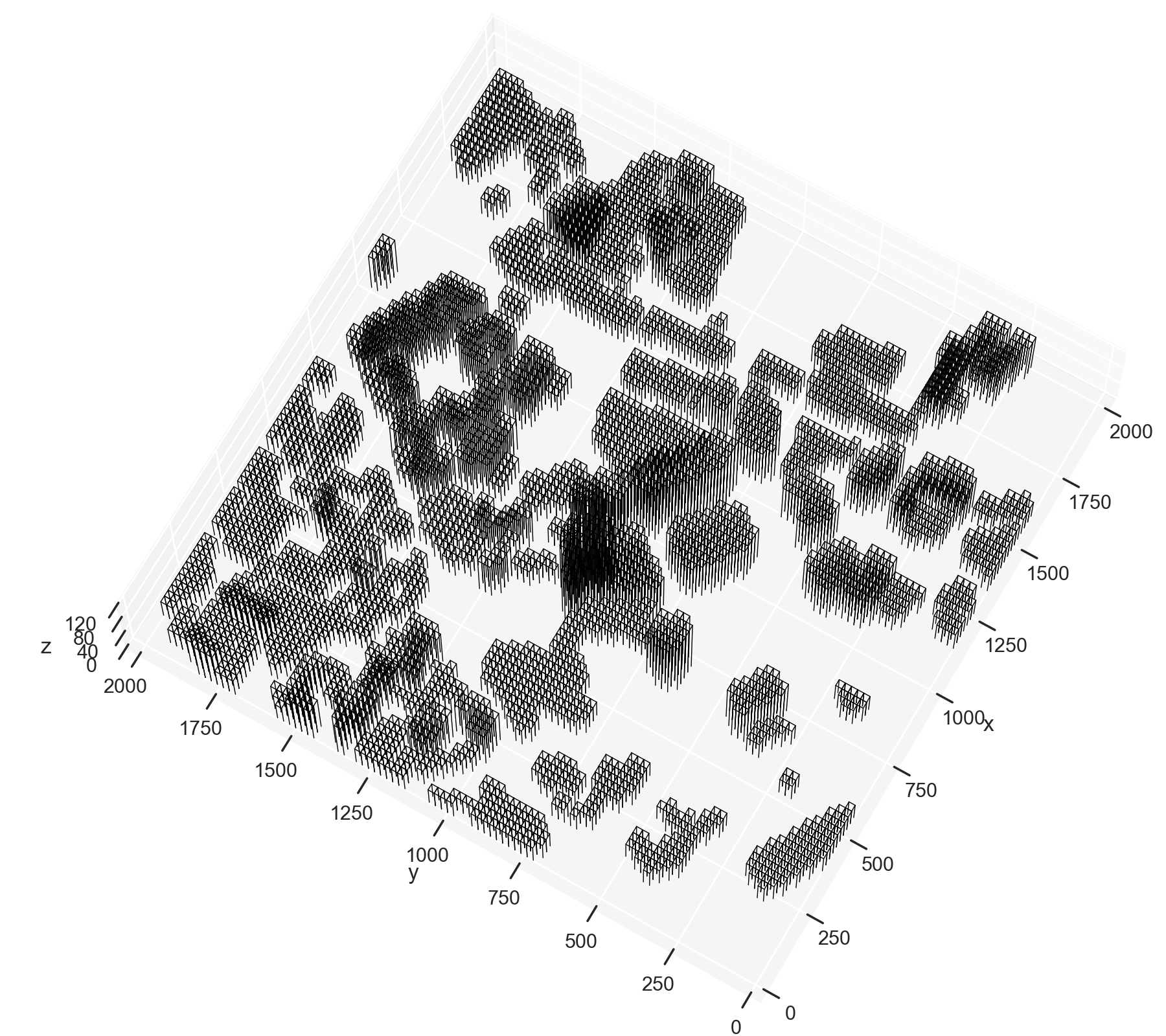}}\label{building} 
          \caption{Scenario for numerical experiment}
        	\label{scenario}
        	\vspace{0.2in}
        \end{figure*}

    The selected area for numerical study is a region in Jurong East, a typical urban area in Singapore, as shown in Figure~\ref{scenario}(a). The size of this area is 2 km $\times$ 2 km. Building information in this area was provided by Singapore Land Authority (SLA). In the building dataset, every building is described by its maximum height, and horizontal outline in terms of a polygon. They cover 3,925 blocks in the bottom layer, 1,286 blocks in the middle layer and 189 blocks in the top layer. The blocks occupied by buildings are illustrated in Figure~\ref{scenario}(b). 
    
    A traffic plan for 300 flights is formed by randomly generated OD pairs within the AirMatrix network, and the aircraft type for each flight is selected from Table~\ref{aircrafts}. We used 60\% of the maximum speed in the planning, which ensures that the 4D path is flyable, and lead to a more energy-efficient setting. The departure times are randomly generated within a time window of 5 minutes. 
        
    
    \begin{table}[h]
        \begin{center}
        \caption{Parameters of Aircraft Types Used in Simulations}
        \label{aircrafts}
        \begin{tabular}{cccccc}
        \hline
        \textbf{Aircraft}   & \textbf{$m$} & \textbf{$v_{m,v}$} & \textbf{$v_{m,h}$} & \textbf{$v_{\beta ,c}$} & \textbf{$v_{\alpha ,c}$}\\ \hline
        DJI Mavic Air       & 0.43 kg       & 4 m/s          & 19 m/s   & 4.454 m/s & 4.860 m/s\\
        Self-Built Drone    & 0.3 kg        & 4 m/s          & 12 m/s   & 4.295 m/s & 4.645 m/s\\
        DJI Phantom 4       & 1.375 kg      & 3 m/s          & 20 m/s   & 3.350 m/s & 3.664 m/s\\
        DJI Matrice 600 Pro & 10 kg         & 5 m/s          & 18 m/s   & 5.713 m/s & 6.258 m/s\\\hline
        \end{tabular}
        \end{center}
        
    \end{table}
    
    \subsection{Results and Discussions}
\begin{figure}[h]
            \centering\includegraphics[width=0.6\linewidth]{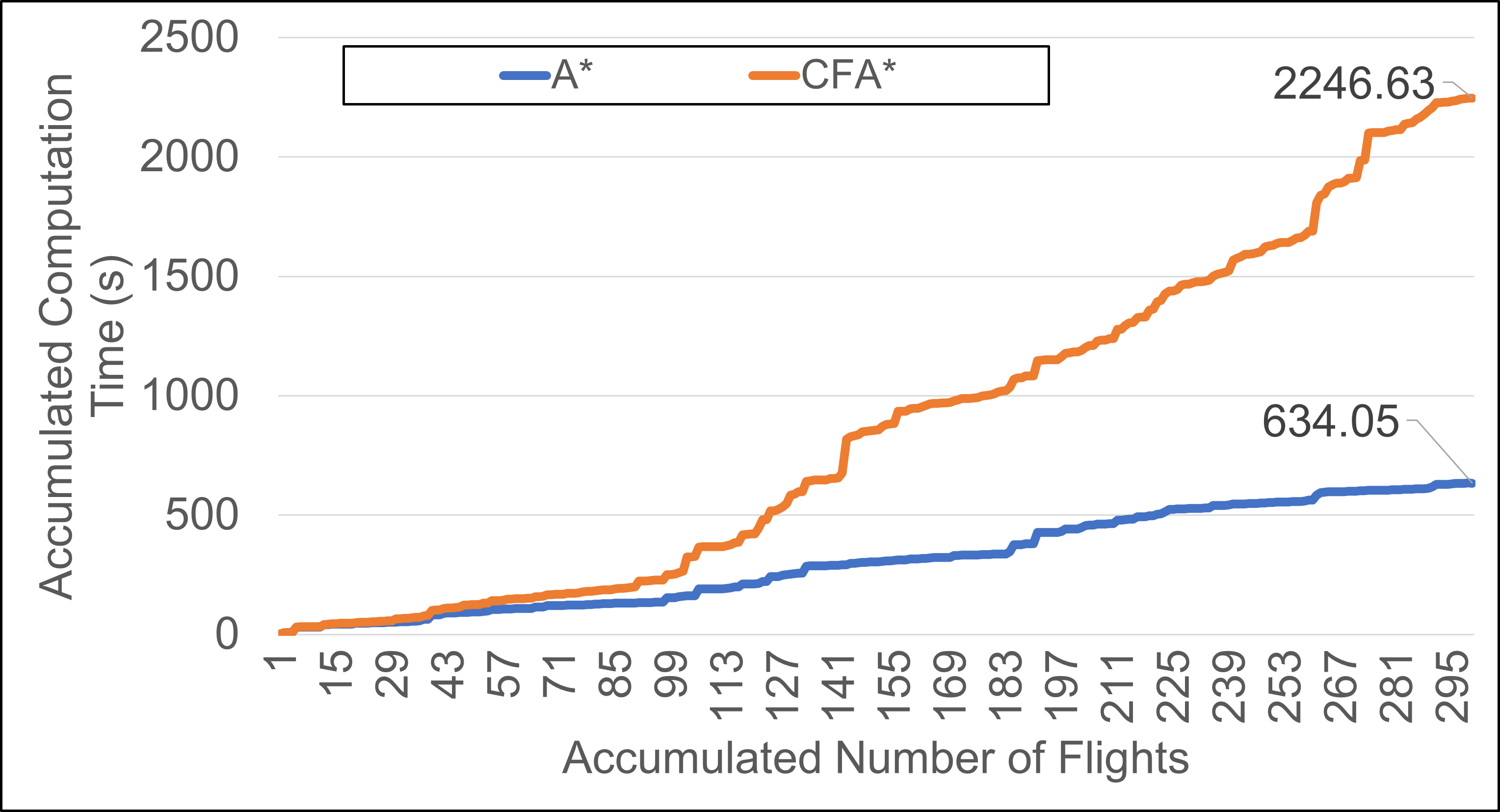}
            \caption{Accumulated computation time of A* and CFA* algorithm}
            \label{accu_comp_time}
        \end{figure}

    The algorithms run on a desktop computer with Intel(R) Core(TM) i9-9900X 3.5GHz CPU. {A* algorithm and CFA* algorithm separately generate routes for the same flight plans.} The computation time for A* and CFA* algorithm is illustrated in Figure~\ref{accu_comp_time}. A* algorithm finished the planning for 300 flights in 634.05 seconds, while the total computation time of CFA* algorithm is 2246.63 seconds. With the increase of existing trajectories, regarded as dynamic obstacles, the computational time of CFA* algorithm for each flight raises, leading to a growing trend of the curve referring to the accumulated computation time of CFA* algorithm. On average, the CFA* algorithm spends 5.4 seconds more than A* algorithm for each flight. Considering the algorithms are meant for pre-flight path planning, this computing cost is acceptable and CFA* is sufficient in terms of computational time.
    
\begin{figure}[]
            \centering\includegraphics[width=0.6\linewidth]{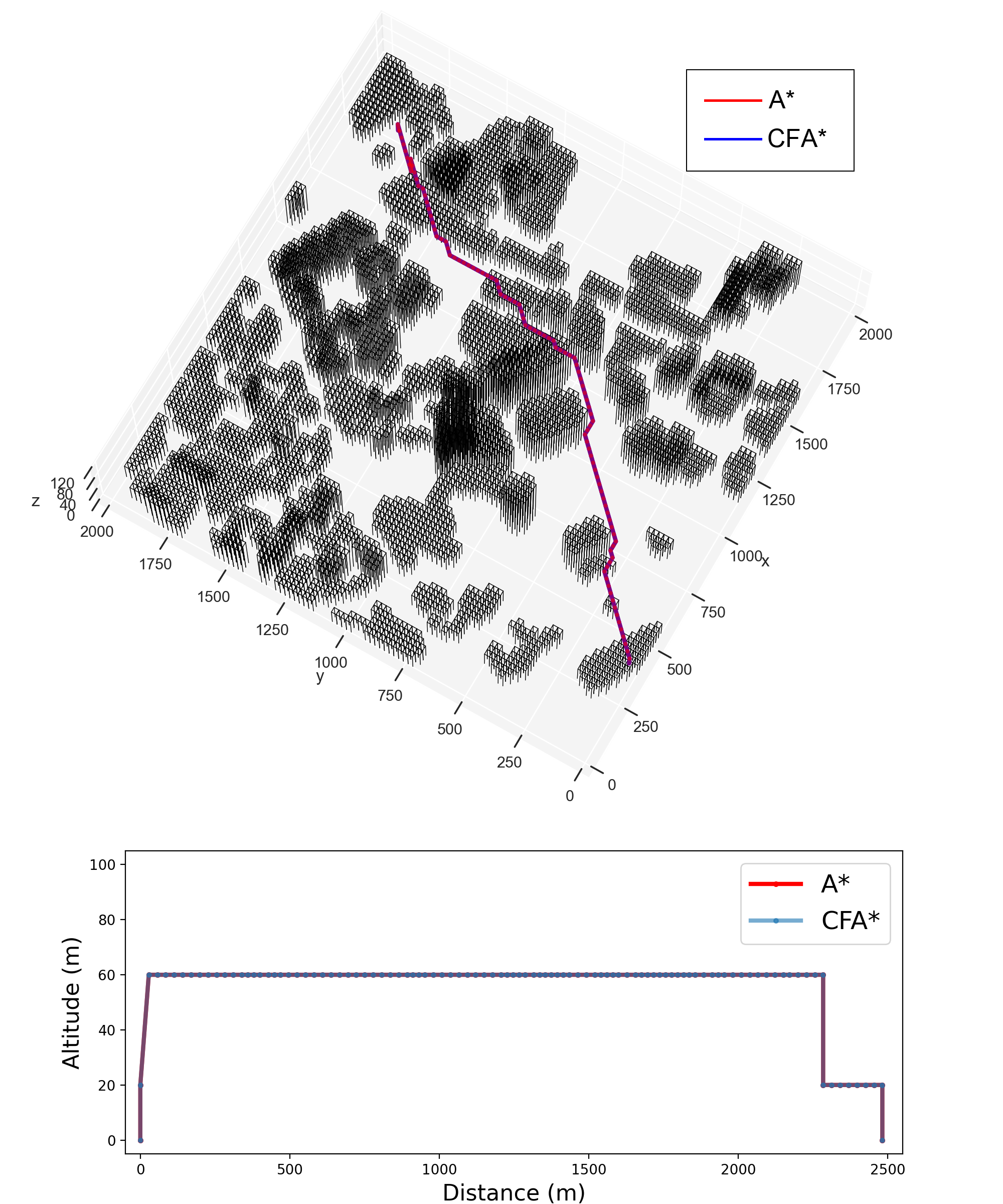}
            \caption{A typical case of the comparison between the trajectories planned by A* and CFA* algorithm: overlapping trajectories}
            \label{overlap}
        \end{figure}

\begin{figure}[]
            \centering\includegraphics[width=0.6\linewidth]{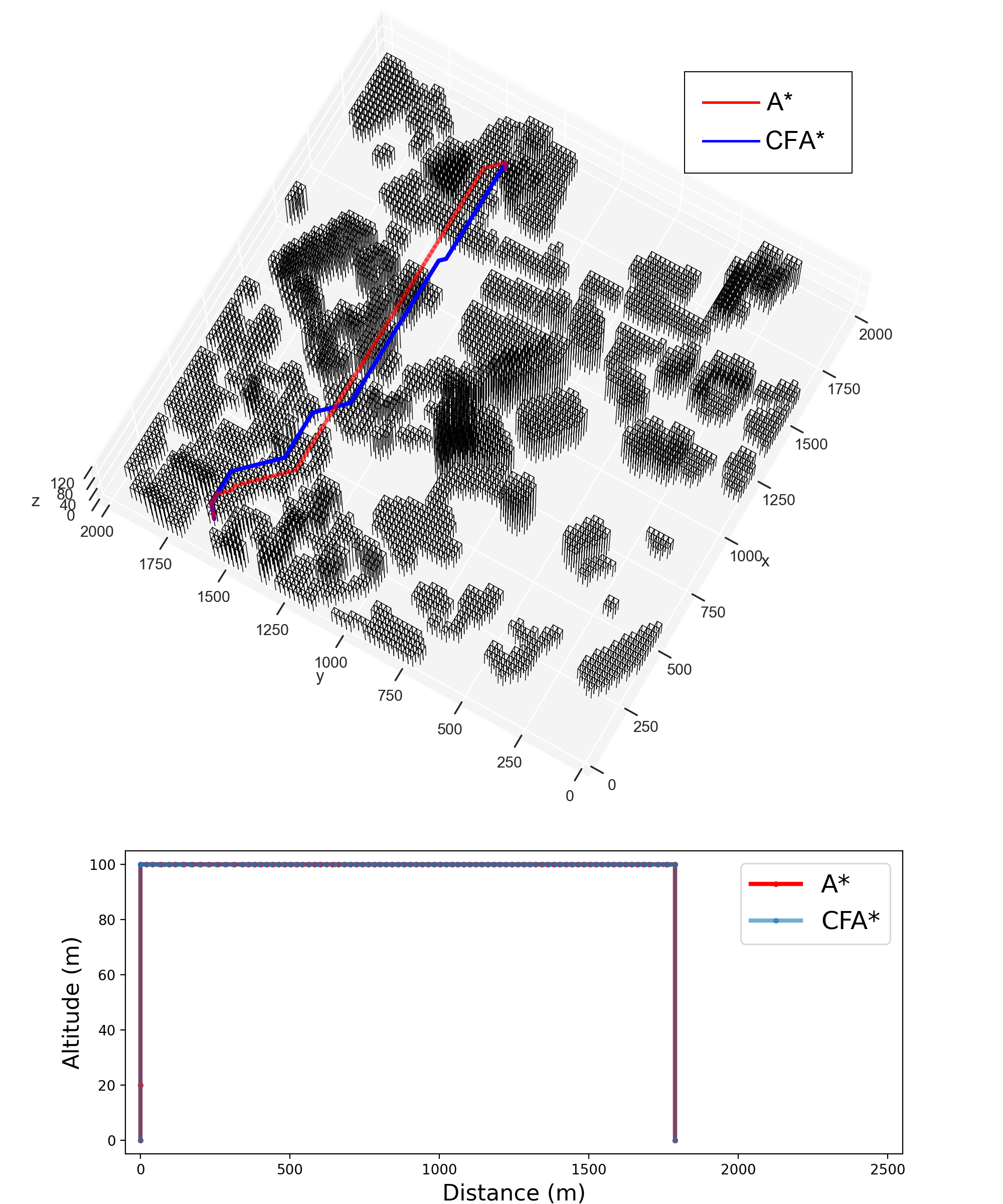}
            \caption{A typical case of the comparison between the trajectories planned by A* and CFA* algorithm: slightly different trajectories with same flight times}
            \label{para}
        \end{figure}

\begin{figure}[]
            \centering\includegraphics[width=0.6\linewidth]{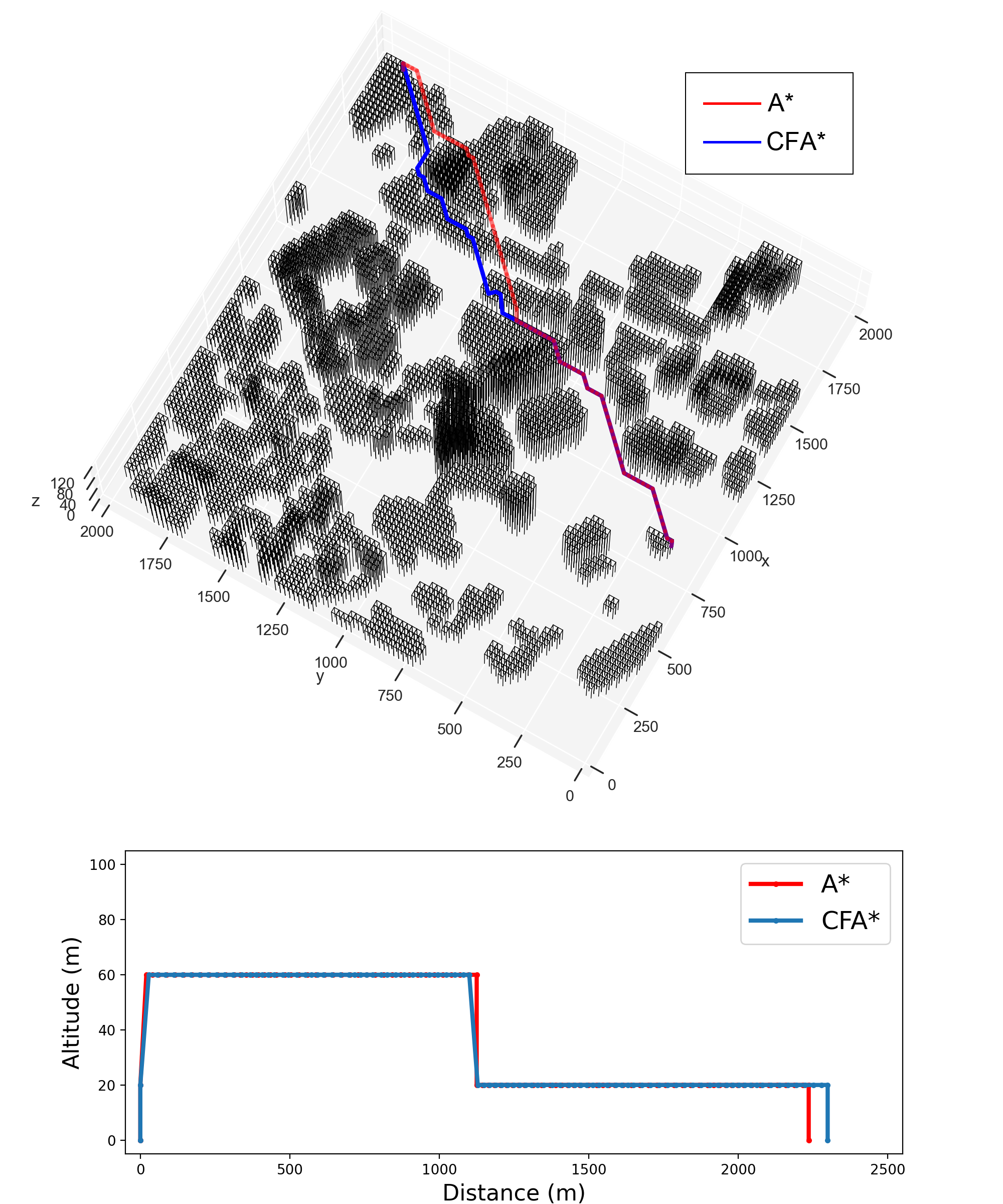}
            \caption{A typical case of the comparison between the trajectories planned by A* and CFA* algorithm: similar trajectories while detouring for conflict avoidance is performed by CFA* algorithm}
            \label{simi}
        \end{figure}

\begin{figure}[]
            \centering\includegraphics[width=0.6\linewidth]{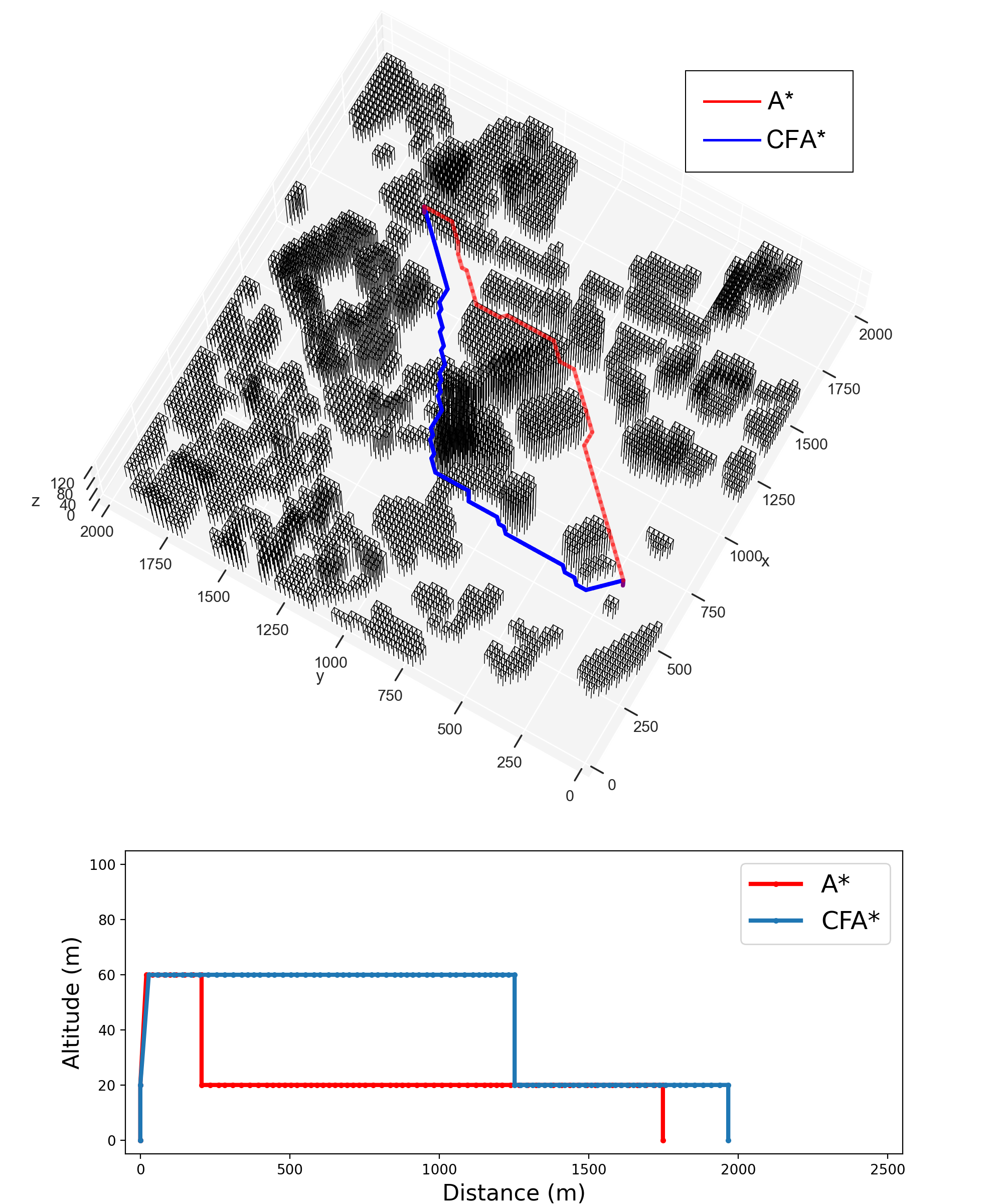}
            \caption{A typical case of the comparison between the trajectories planned by A* and CFA* algorithm: totally different trajectories}
            \label{total_diff}
        \end{figure}

    {Four typical cases of the trajectories planned by A* and CFA* algorithms as well as illustrations of their vertical profiles are shown in Figure~\ref{overlap} to~\ref{total_diff}. Figure~\ref{overlap} shows that no conflict is detected and CFA* algorithm outputs an optimal trajectory which is exactly the same to trajectory planned by A* algorithm. In the case illustrated by Figure~\ref{para}, the two trajectories are slightly different, but they have the same flight time. Figure~\ref{simi} illustrates a typical scenario that CFA* detours due to conflicts of block utilization is detected. In Figure~\ref{total_diff}, the trajectories planned by A* and CFA* algorithm are totally different. The differences between the trajectory planned by A* algorithm and CFA* algorithm shown in Figure~\ref{para} to~\ref{total_diff}are because of the conflicts detected during path planning, and CFA* performed detouring as a solution. Among them, the conflict avoidance performed by CFA* in the trajectory shown in Figure\ref{pra} doesn't cause any flight delay, indicating there's no hovering required, and when the route planned by A* algorithm would cause a conflict, the algorithm found an alternate route. }
    
        
\begin{figure}[htb]
            \centering\includegraphics[width=0.9\linewidth]{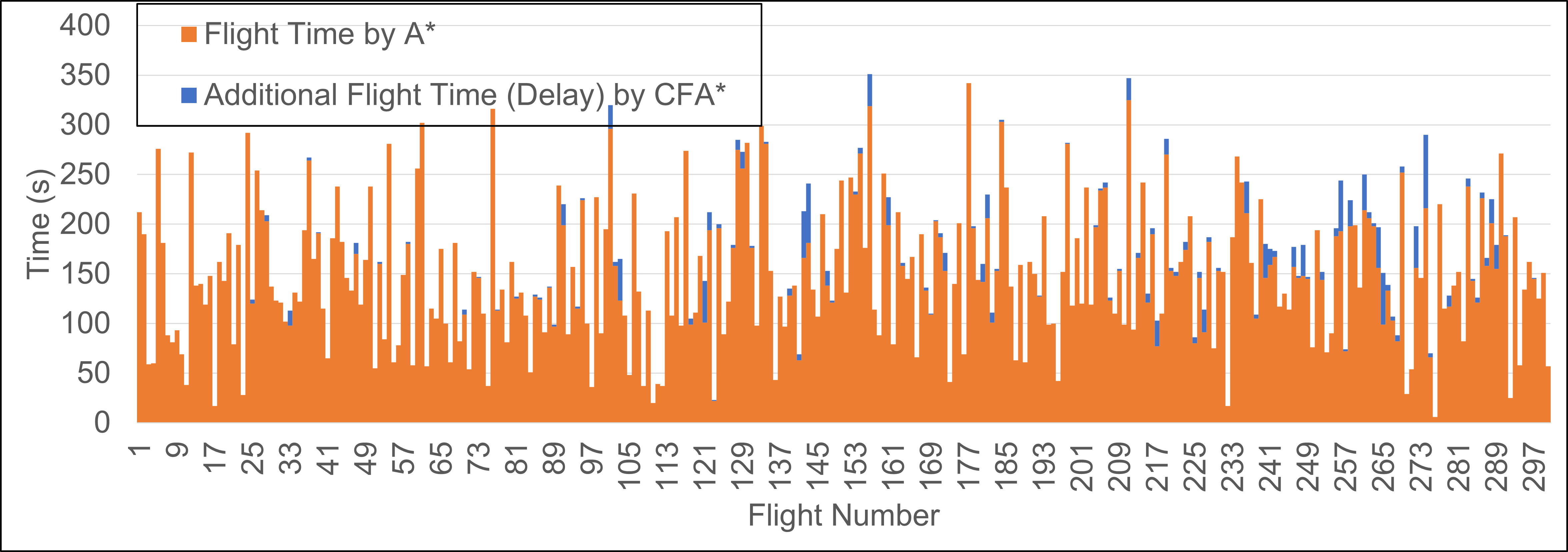}
            \caption{Flight delay of the trajectories planned by CFA* compared with A* algorithm}
            \label{cost_and_delay}
        \end{figure}

\begin{figure}[htb]
            \centering\includegraphics[width=0.9\linewidth]{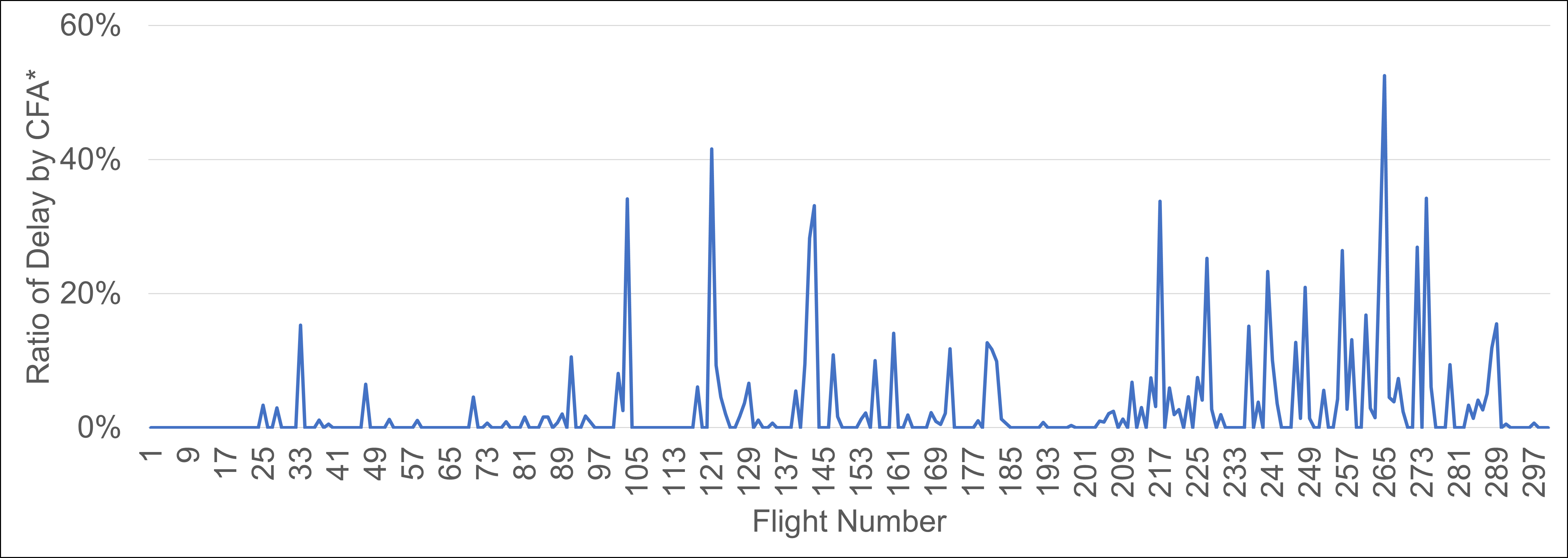}
            \caption{Ratio of flight delay of the trajectories planned by CFA* compared with A* algorithm}
            \label{delay_ratio}
        \end{figure}

    Figure~\ref{cost_and_delay} illustrates the differences between the flight time of the trajectories planned by A* algorithm and CFA* algorithm, namely flight delay. The orange bars refer to the flight time planned by A* algorithm, and the blue bars refer to the flight delay. Therefore the total height of each bar denotes the flight time of the trajectory planned by CFA* algorithm. The maximum flight time of the 300 flights is 351 seconds. {When the accumulated number of flights is relatively small, conflict avoidance is less required and most of the trajectories planned by CFA* algorithm has the same performance in terms of flight time compared with A* algorithm, as shown in the left part of this figure that there are only a few flight delay. While the accumulated number of flights increases, avoidance is required from existing planned trajectories. So that more flight delay can be observed in the right part of the figure. Another illustration of this phenomenon is illustrated in Figure~\ref{delay_ratio}, which shows the delay ratio of the trajectories planned by CFA* compared with A*. }
        
\begin{figure}[htb]
            \centering\includegraphics[width=0.9\linewidth]{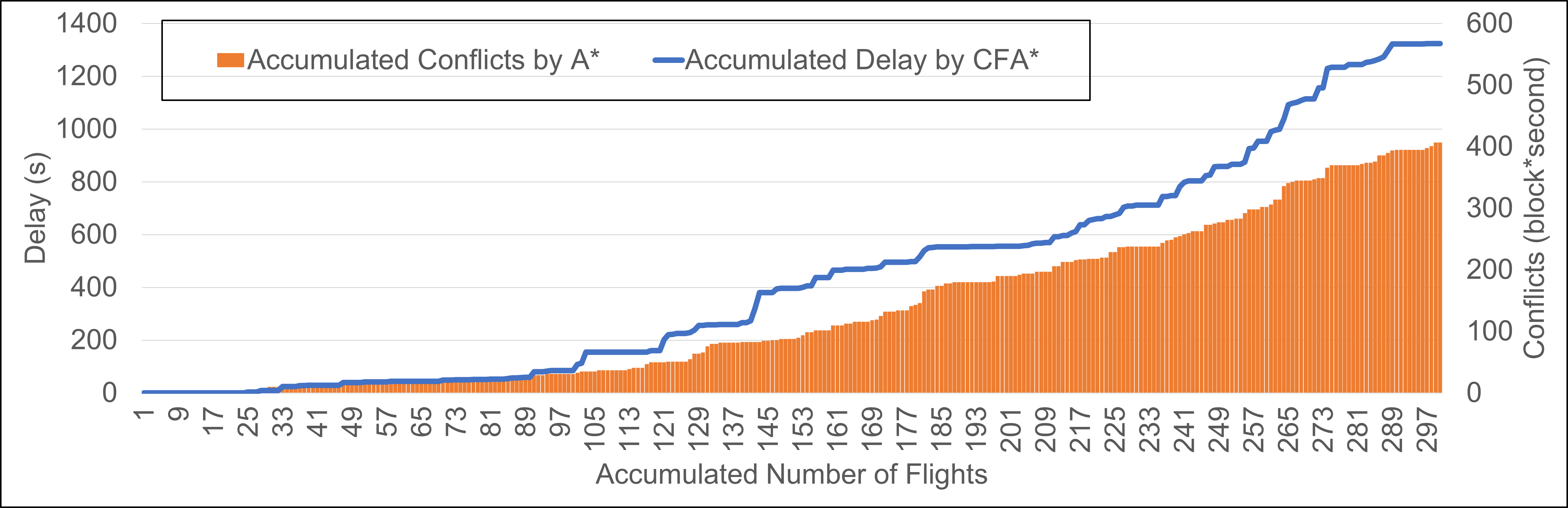}
            \caption{Accumulated delay time of the trajectories planned by CFA*}
            \label{accu_delay}
        \end{figure}

    Figure~\ref{accu_delay} illustrates the accumulated delay by CFA* algorithm and the accumulated number of conflicts by A* algorithm, {where the conflict is quantified as the sum of the time each block is duplicated occupied:}
    \begin{equation}
        {\sum_{i,j,k,t} U(\sum_{Path}Occup(Path)_{i,j,k,t}-1),\ Path \in \{all\ planned\ paths\}}
    \end{equation}
    \begin{equation}
        {U(x) = \left\{ \begin{array}{ll}
            1, & x>0\\
            0, & x\leq 0
            \end{array}
            \right.}
    \end{equation}
    Due to the high flight density, both accumulations show very high values at the right end. Compared with A* algorithm, the CFA* algorithm resolved 407 conflicts in block utilization, which has a significant impact on flight safety. The two curves shown in this figure has a very similar growing trend with the increasing number of flights.

\begin{figure}[htb]
            \centering\includegraphics[width=0.9\linewidth]{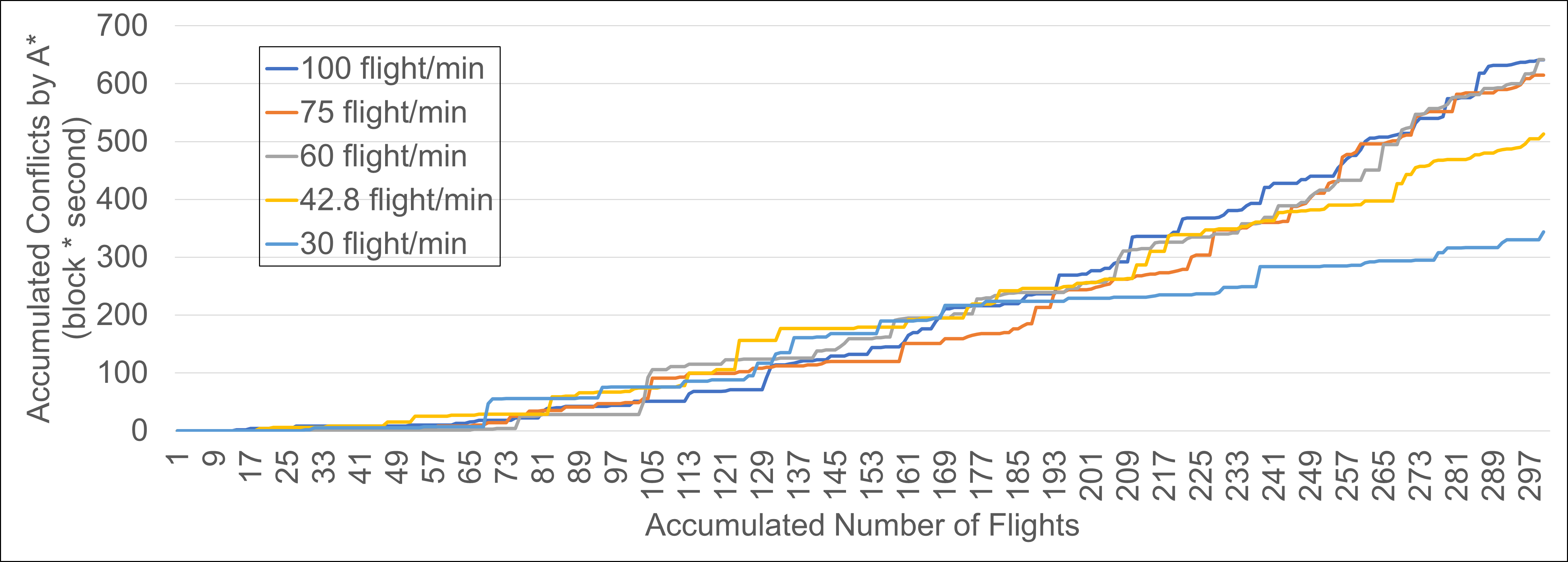}
            \caption{Accumulated conflicts of the trajectories planned by A* in the cases with different flight density}
            \label{sens_conflict}
        \end{figure}

\begin{figure}[htb]
            \centering\includegraphics[width=0.9\linewidth]{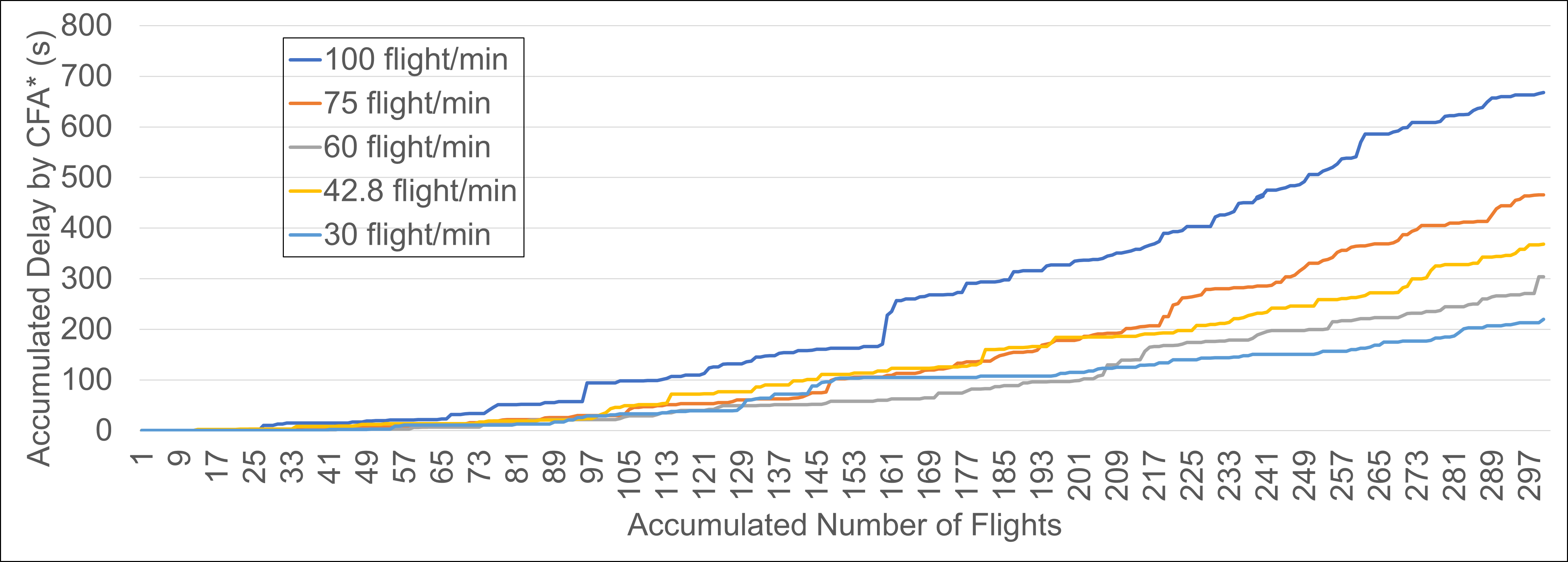}
            \caption{Accumulated delay of the trajectories planned by CFA* in the cases with different flight density}
            \label{sens_delay}
        \end{figure}

    {Sensitivity analysis was carried out to investigate the impact of flight density on the planning performance in terms of the accumulated number of conflicts resulted from the A* algorithm planning, and the accumulated flight delay resulted from CFA* algorithm planning. Five cases with different flight densities were simulated. Due to the FCFS setting of the algorithms, there's little point in changing the number of flight plans. Because the result of simulating 150 flights are the same to simulating 300 flights and selecting the first 150 of them. Hence, we generated 300 flight plans for each case, but modified the time window of departure to change the flight density. The departure time windows of 3, 4, 5, 7 and 10 minutes were selected, leading to 100, 75, 60, 42.8 and 30 flights per minutes, respectively. The results of sensitivity analysis show that, in general, the higher flight density, the larger flight delay is resulted from CFA* algorithm, as shown in Figure~\ref{sens_delay}. Also as shown in Figure~\ref{sens_conflict}, the higher flight density, the larger number of conflicts. However, in Figure~\ref{sens_conflict}, the increasing trends of conflicts for the cases with 100, 75, and 60 flights per minute are similar, while the number of conflicts decreases significantly when the flight density is less than 60 flights per minute. In Figure~\ref{sens_delay}, 60 flights per minute is leading to a smaller delay than 42.8 flight per minute. These are because the flight plans were randomly generated, so that the possibility of trajectory conflict is unpredictable. The results of sensitivity analysis can reveal the trend of the relationship between flight density and the performances of planned paths. But the results should be viewed qualitatively, not quantitatively. }
    

\begin{figure*}[]
          \centering
            \subfigure[Bottom layer]{\includegraphics[width=0.3\textwidth]{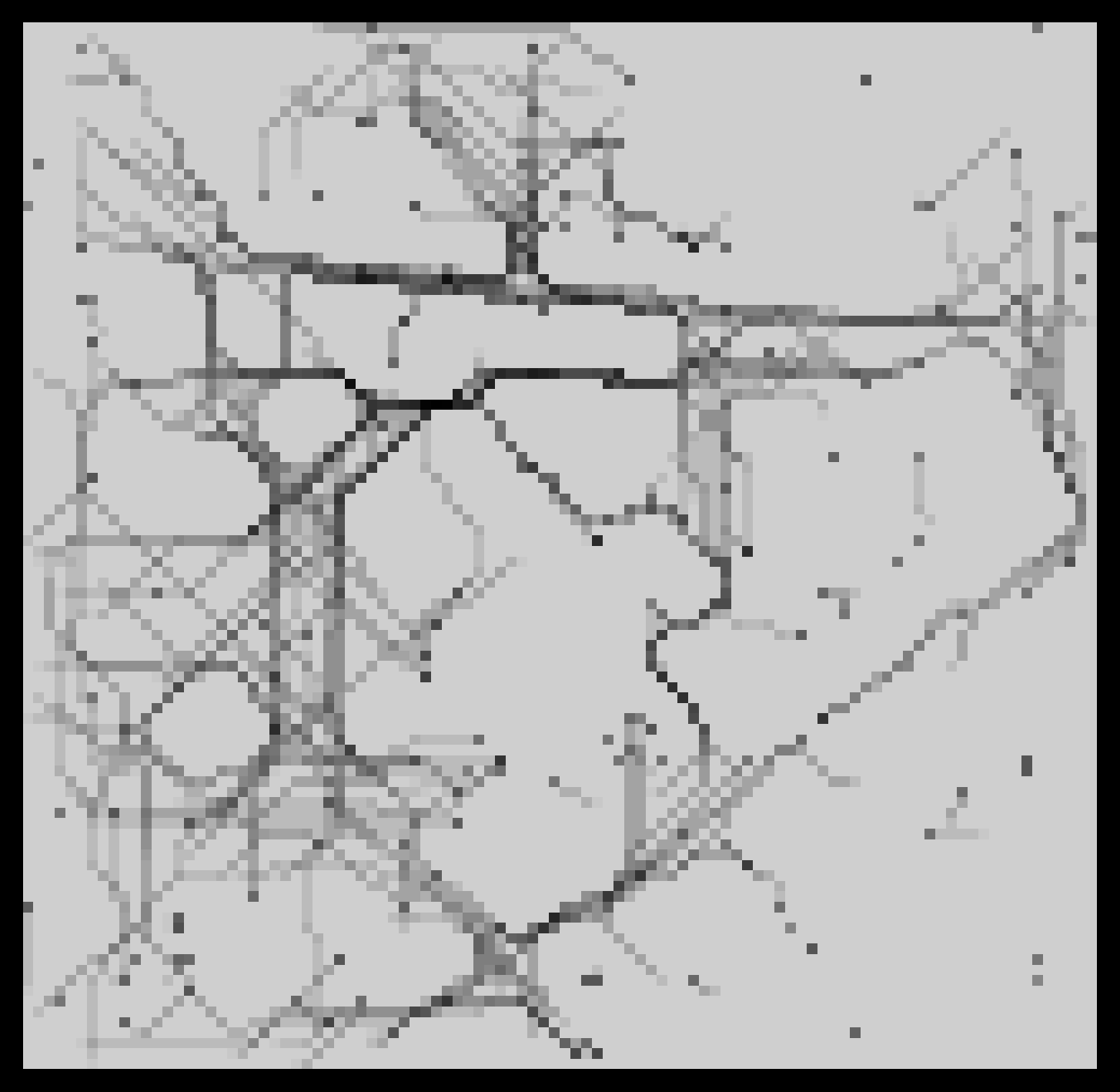}} 
        	\subfigure[Middle layer]{\includegraphics[width=0.3\textwidth]{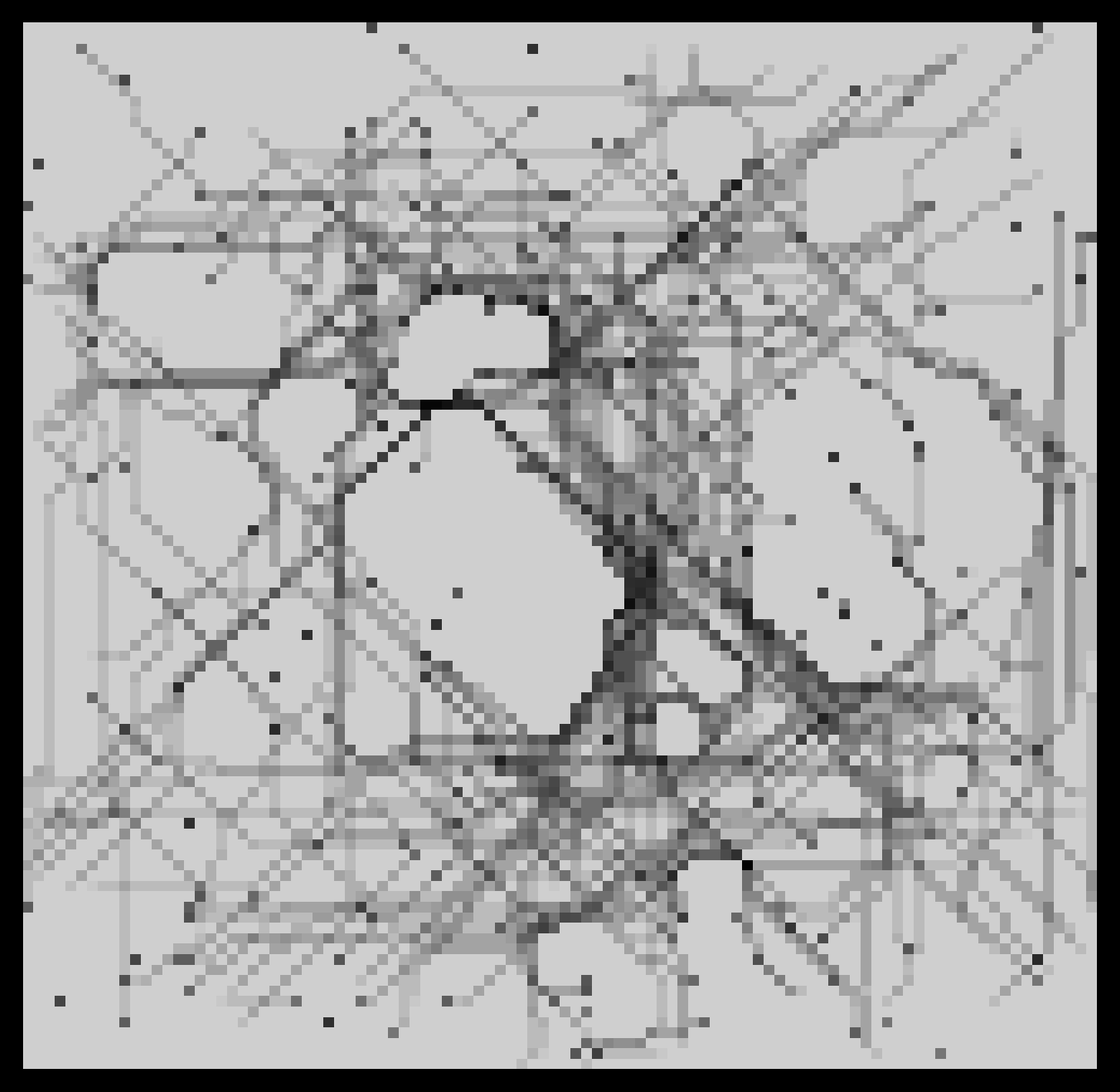}} 
            \subfigure[Top layer]{\includegraphics[width=0.3\textwidth]{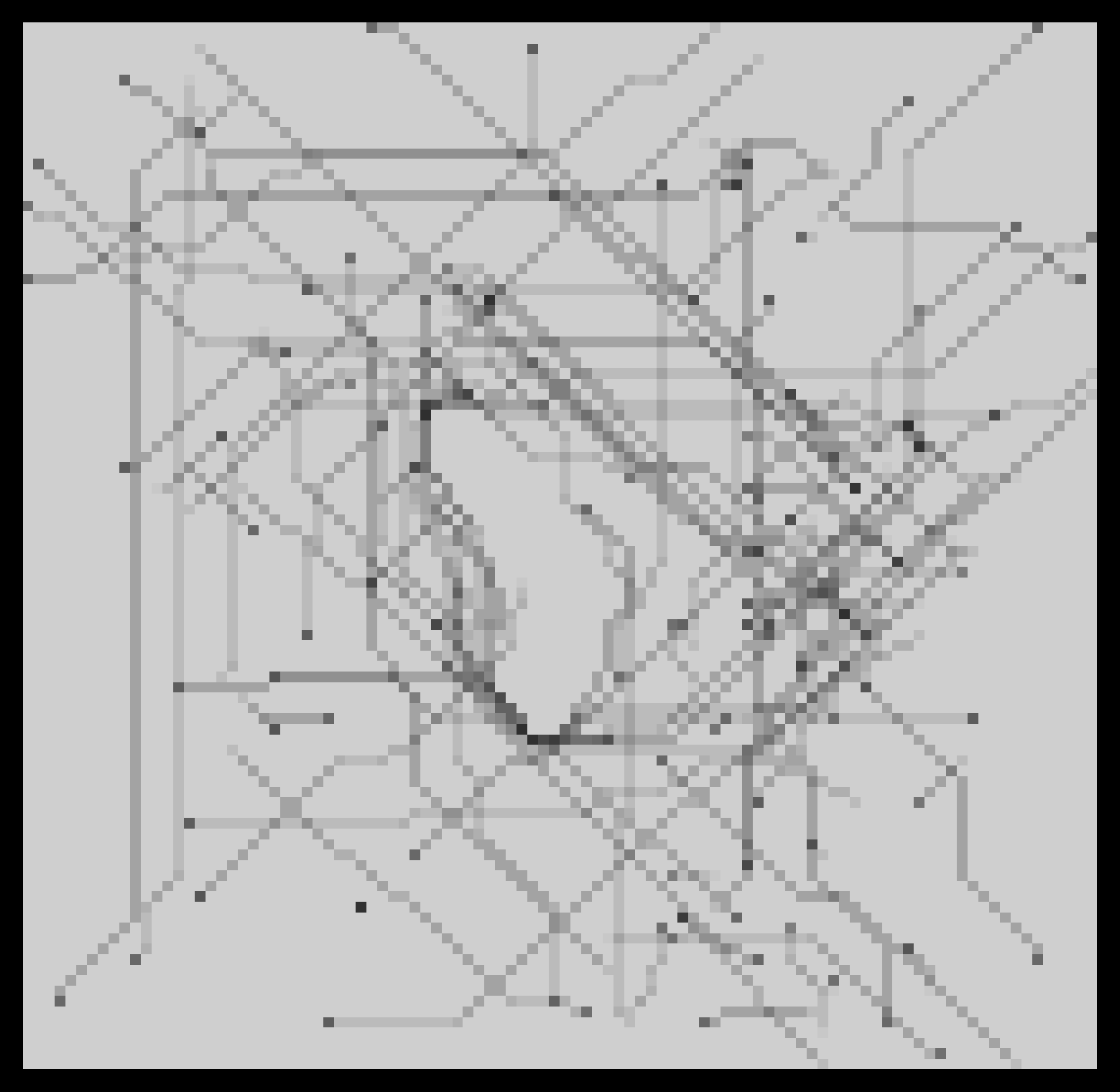}} 
          \caption{Heatmap of airspace utilization in the three layers of AirMatrix blocks}
        	\label{heatmap}
        	\vspace{0.2in}
        \end{figure*}

        Figure~\ref{heatmap} shows the quantification of airspace utilization in the three layers. The utilization of the bottom layer is illustrated in Figure~\ref{heatmap}(a), where the darker the colour of each block, the more the block is utilized. Considering almost 40\% of the blocks in this layer are occupied by buildings, there's not too much room for aircraft to fly. But we can see a clear pattern of several corridors that are frequently used. Figure~\ref{heatmap}(b) shows the airspace utilization of the middle layer. Buildings occupy only 12.9\% of the blocks in this layer, which is much less than the bottom layer. Therefore this layer is mostly used. Figure~\ref{heatmap} (c) shows the airspace utilization of the top layer. Considering the vertical velocity of aircraft is small, thus climbing to a high altitude requires a relatively long time, which is against the preference in the algorithm. This leads to a low utilization rate of high-altitude airspace. However, from the perspective of airspace management, the use of airspace at different altitudes should be relatively even. Because spreading flights evenly to various altitudes helps to maintain safe separation and avoid congestion. The method of trajectory planning with a consideration of altitude assignment should be studied in the future.
        
        {The objective function to be minimized in CFA* is flight time. Therefore the use of different layers of airspace depends on which time cost is smaller: maneuvering in the horizontal direction or in the vertical direction. When the sizing of each block is static, the relationship between the horizontal and vertical velocity influences the use of airspace layers. To verify this assumption, we performed three simulations. The flight plans used for the three simulations are the same, but all the flight plans in each simulation are executed by the same type of aircraft: DJI Mavic Air for the first simulation, self-built drone for the second simulation, and DJI Phantom 4 for the third simulation. The airspace layers utilization in the simulation results is shown in Table~\ref{heat_comp}. The ratio of maximum horizontal velocity to maximum vertical velocity for these three aircraft types are 4.75, 3, and 6.66, respectively. It can be found that self-built drone has the highest utilization of the middle layer and the top layer, while DJI Phantom 4 has the lowest. These results show that larger ratio of horizontal velocity to vertical velocity leads to less use of high altitude airspace. }

        It is worth to note that the result of numerical study highly relies on the traffic plan which is randomly generated in this study. The values in the result may vary given different scenarios. However the trend of these results with the accumulation of the number of flights is valid. 
    \begin{table}[]
    \center
        \caption{Airspace Utilization by Different Aircraft Type}
        \label{heat_comp}
    \begin{tabular}{llll}
    \hline
           & \multicolumn{1}{c}{\textbf{DJI Mavic Air}} & \multicolumn{1}{c}{\textbf{Self-Built Drone}} & \multicolumn{1}{c}{\textbf{DJI Phantom 4}} \\ \hline
    Bottom &  \begin{tabular}[c]{@{}l@{}}\begin{minipage}[b]{0.3\columnwidth}\centering \raisebox{-.5\height}{\includegraphics[width=\linewidth]{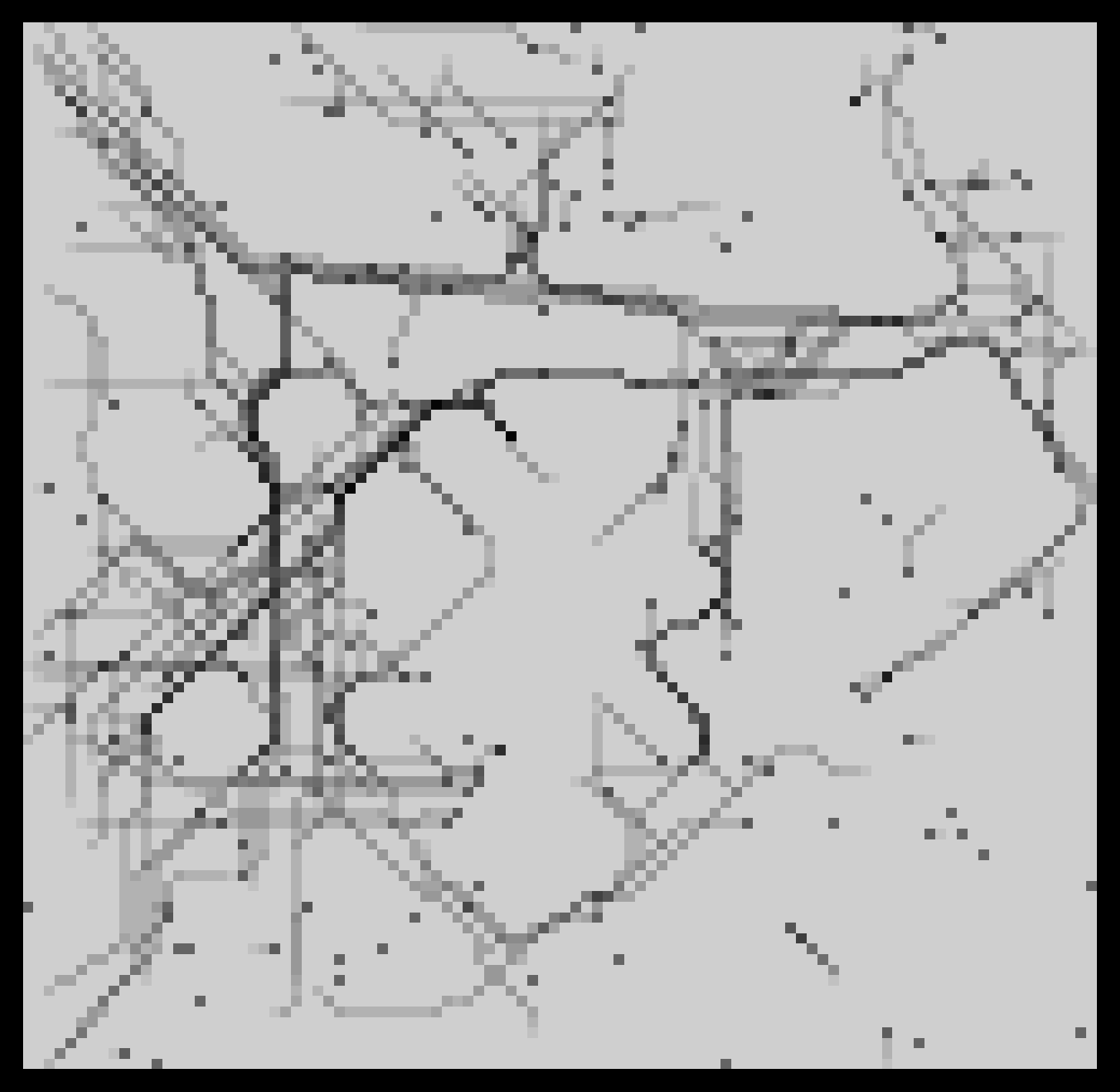}}
	\end{minipage}\\  \end{tabular}  &\begin{tabular}[c]{@{}l@{}}\begin{minipage}[b]{0.3\columnwidth}\centering \raisebox{-.5\height}{\includegraphics[width=\linewidth]{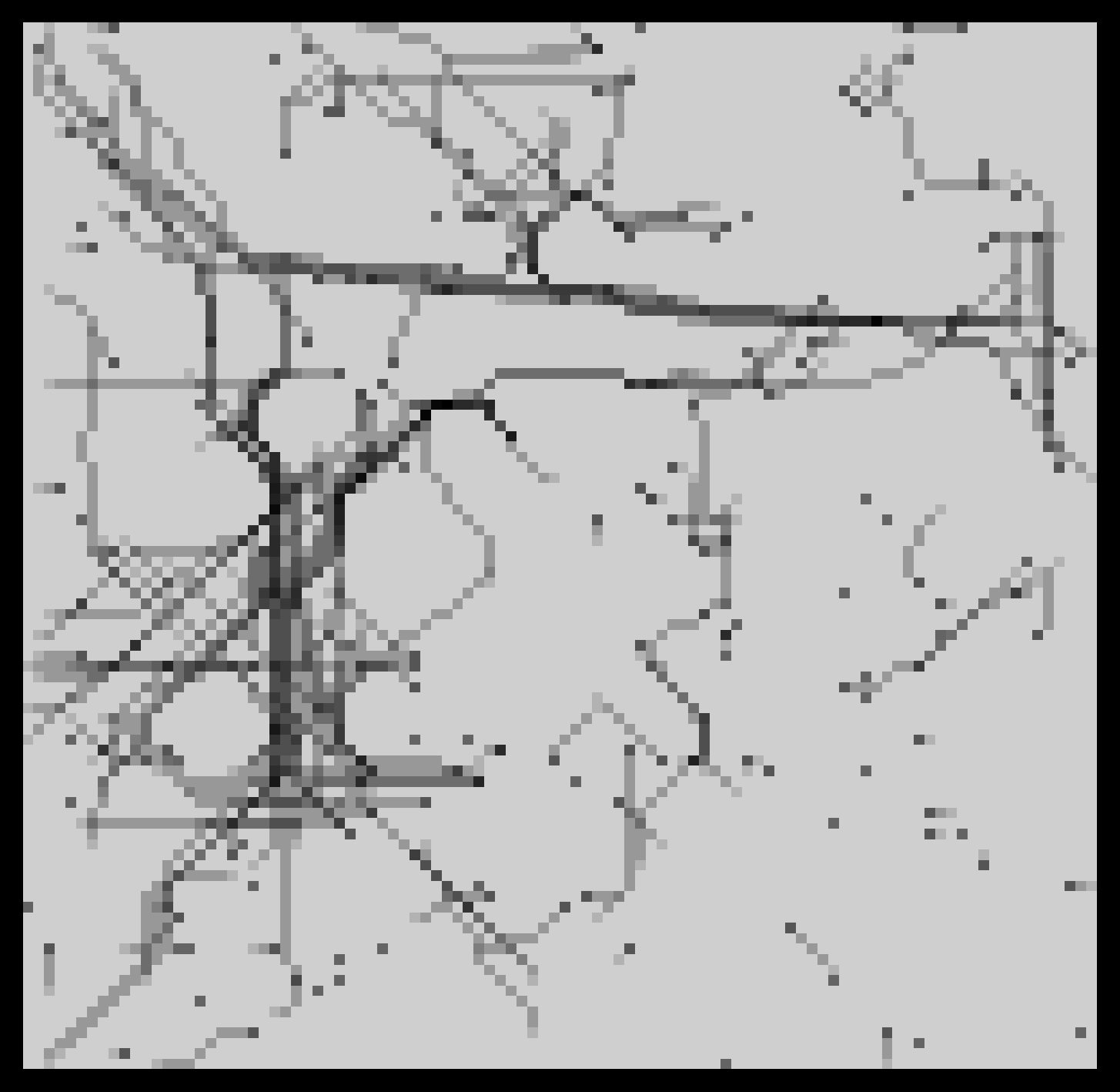}}
	\end{minipage}\\  \end{tabular}&\begin{tabular}[c]{@{}l@{}}\begin{minipage}[b]{0.3\columnwidth}\centering \raisebox{-.5\height}{\includegraphics[width=\linewidth]{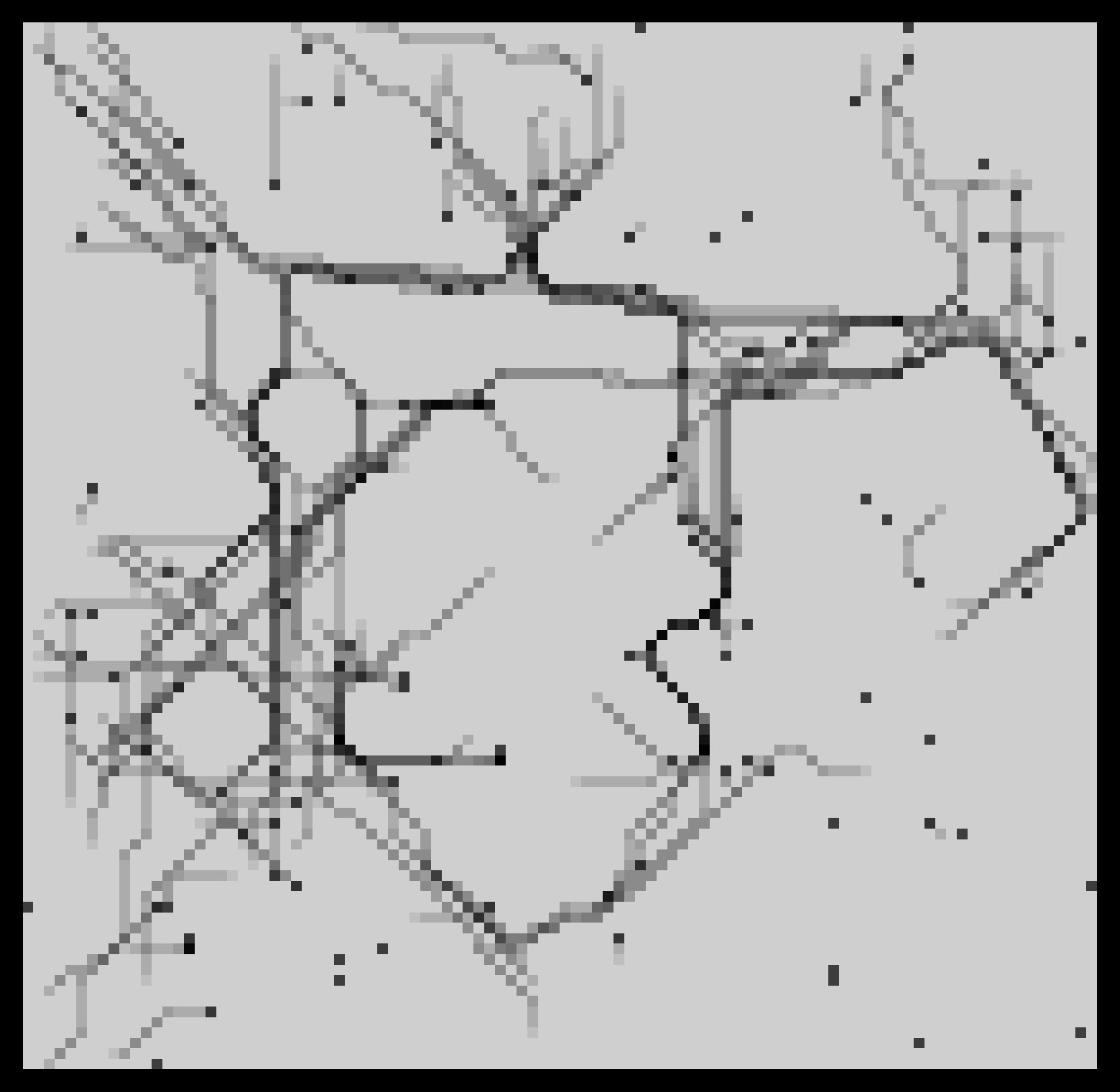}}
	\end{minipage}\\  \end{tabular}\\
           &                                            &                                               &                                            \\
    Middle &\begin{minipage}[b]{0.3\columnwidth}\centering \raisebox{-.5\height}{\includegraphics[width=\linewidth]{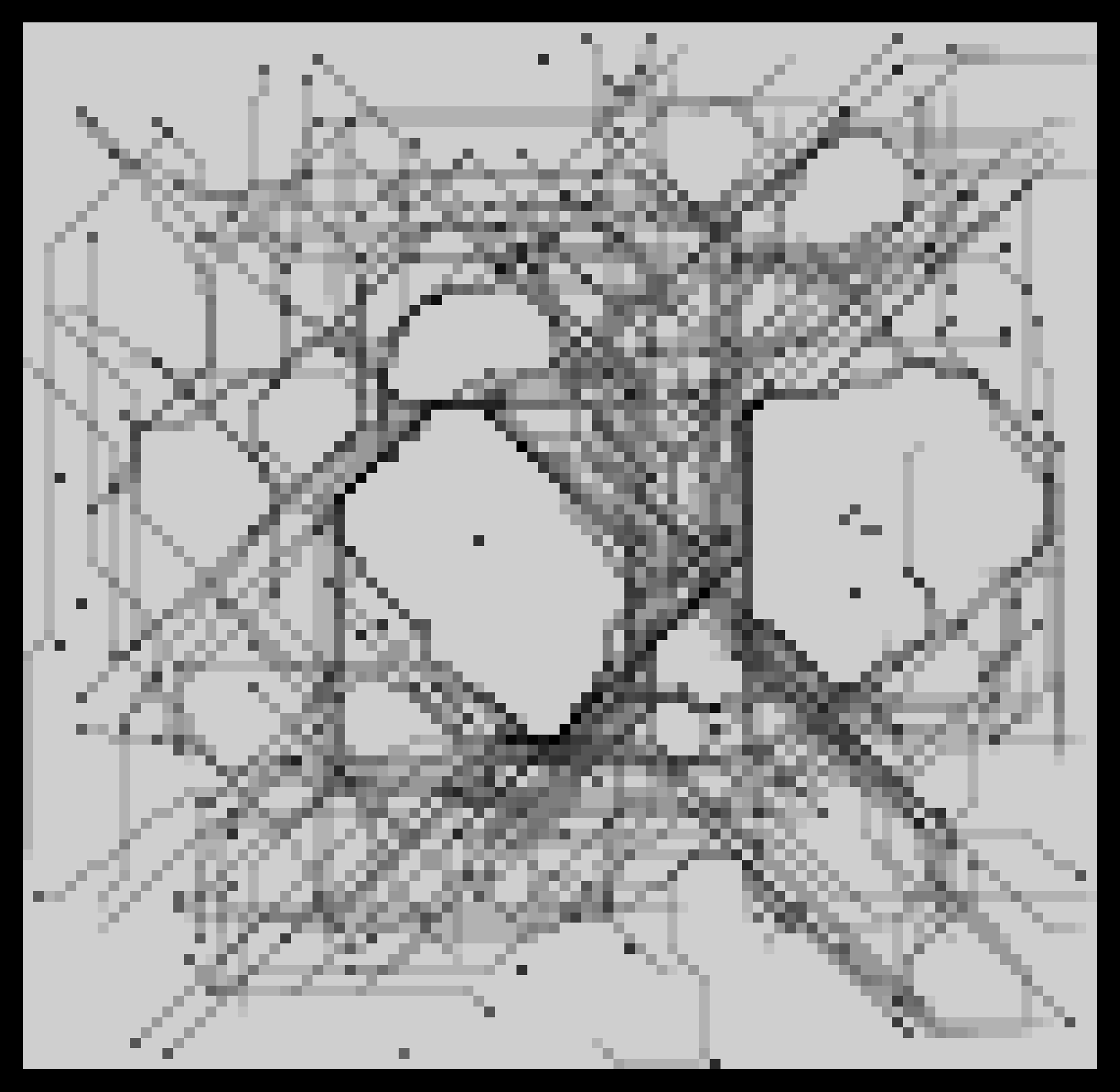}}
	\end{minipage}&\begin{minipage}[b]{0.3\columnwidth}\centering \raisebox{-.5\height}{\includegraphics[width=\linewidth]{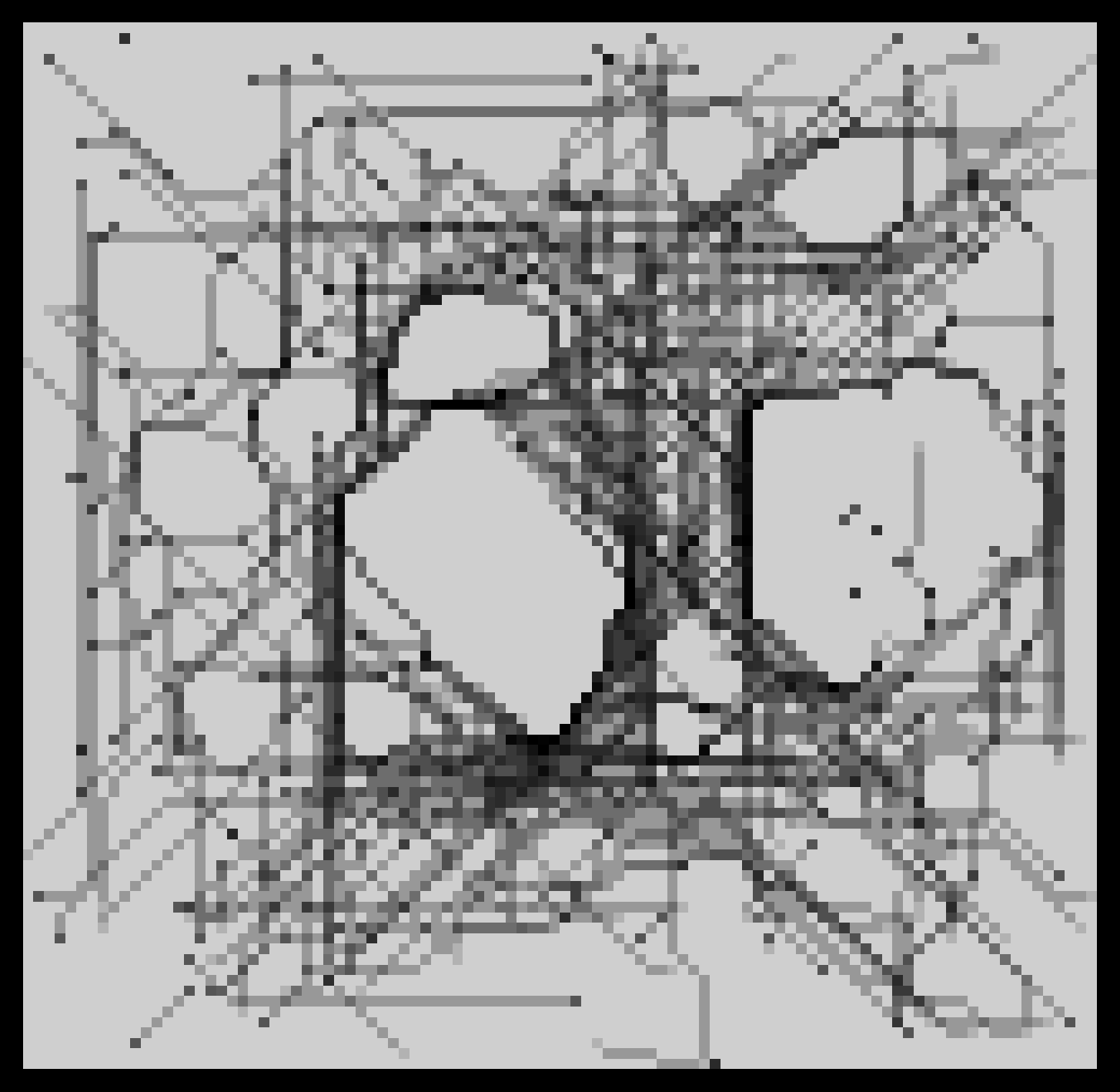}}
	\end{minipage}&\begin{minipage}[b]{0.3\columnwidth}\centering \raisebox{-.5\height}{\includegraphics[width=\linewidth]{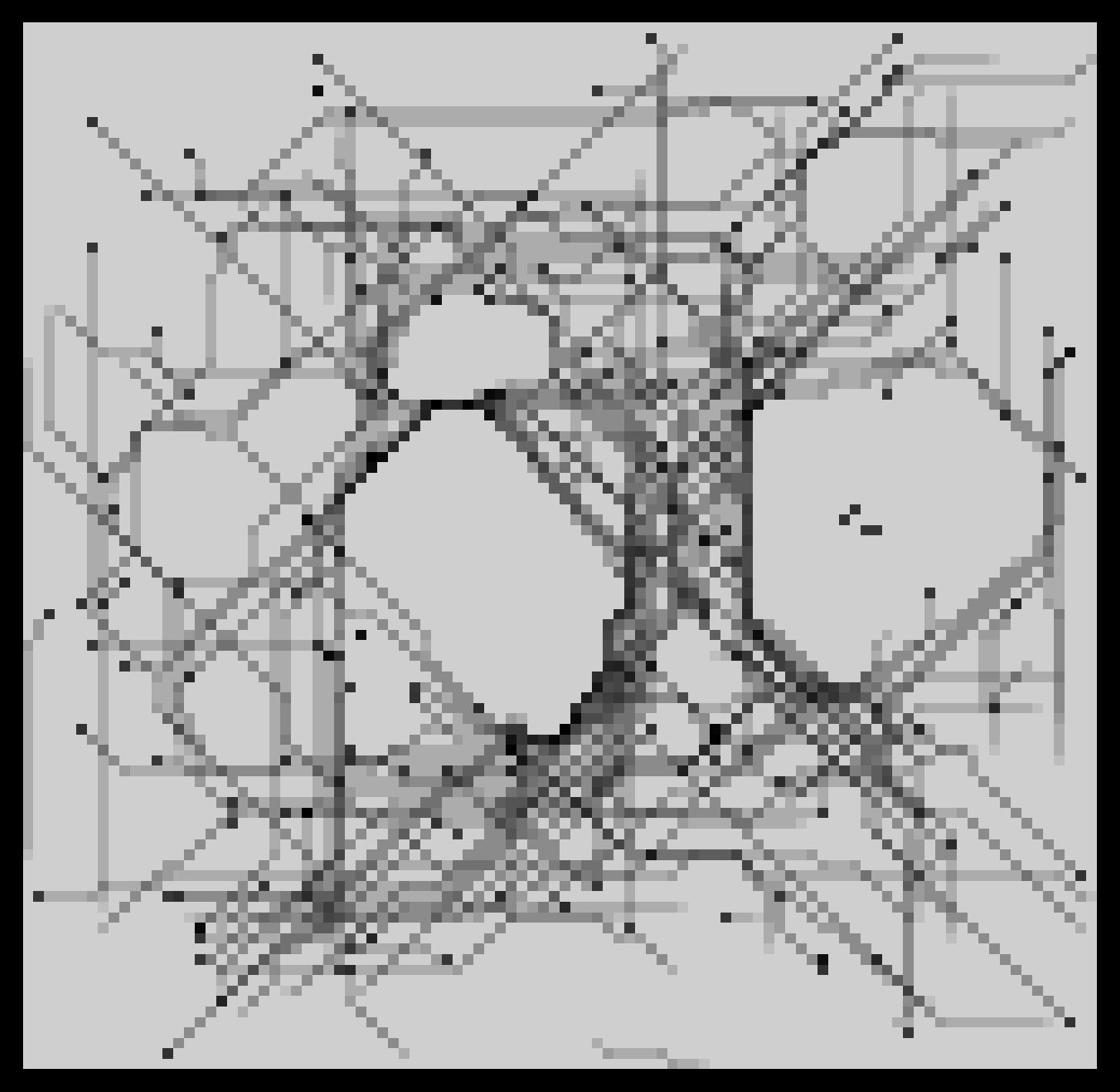}}
	\end{minipage}\\
           &                                            &                                               &                                            \\
    Top    &\begin{minipage}[b]{0.3\columnwidth}\centering \raisebox{-.5\height}{\includegraphics[width=\linewidth]{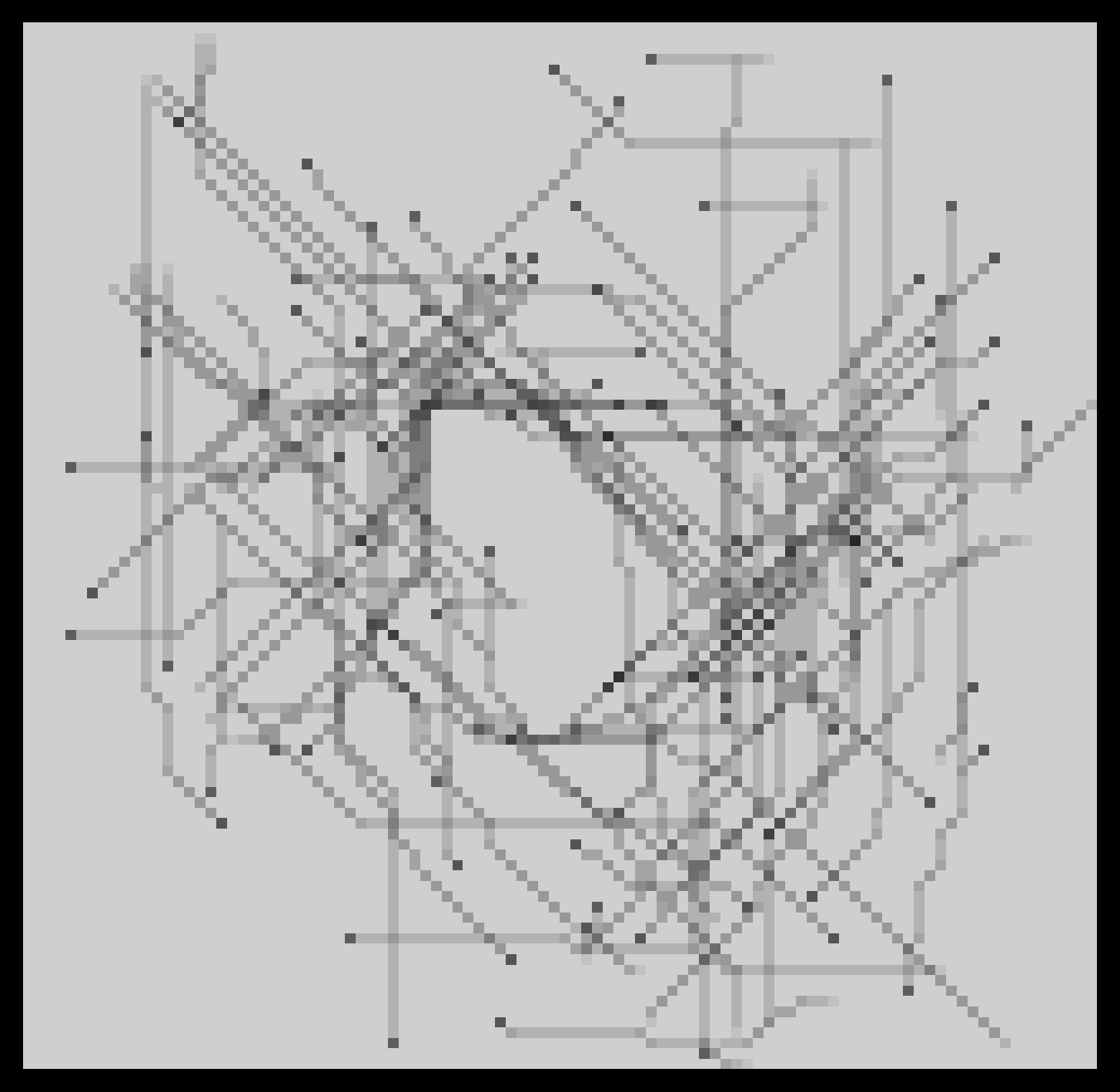}}
	\end{minipage}&\begin{minipage}[b]{0.3\columnwidth}\centering \raisebox{-.5\height}{\includegraphics[width=\linewidth]{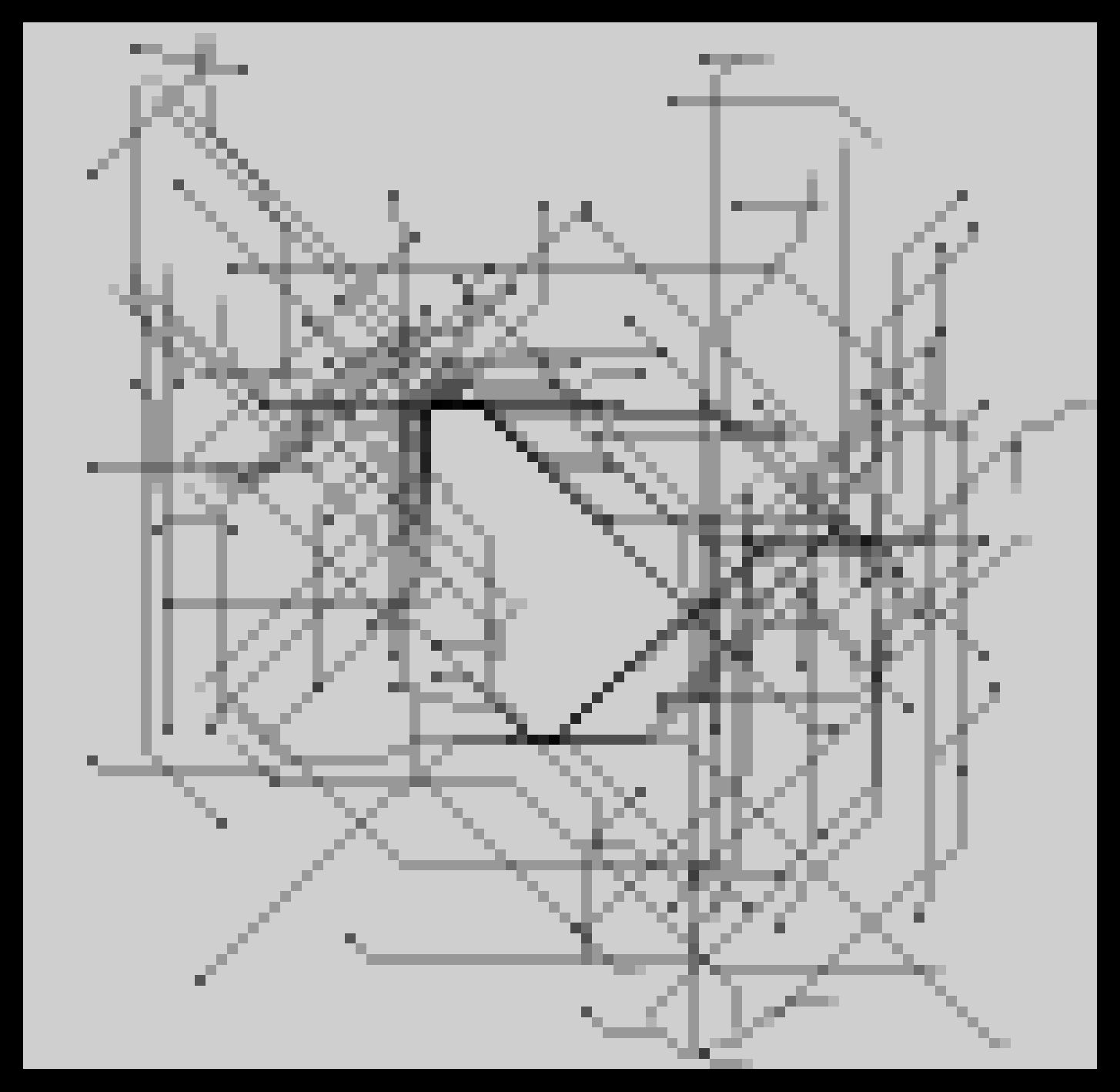}}
	\end{minipage}&\begin{minipage}[b]{0.3\columnwidth}\centering \raisebox{-.5\height}{\includegraphics[width=\linewidth]{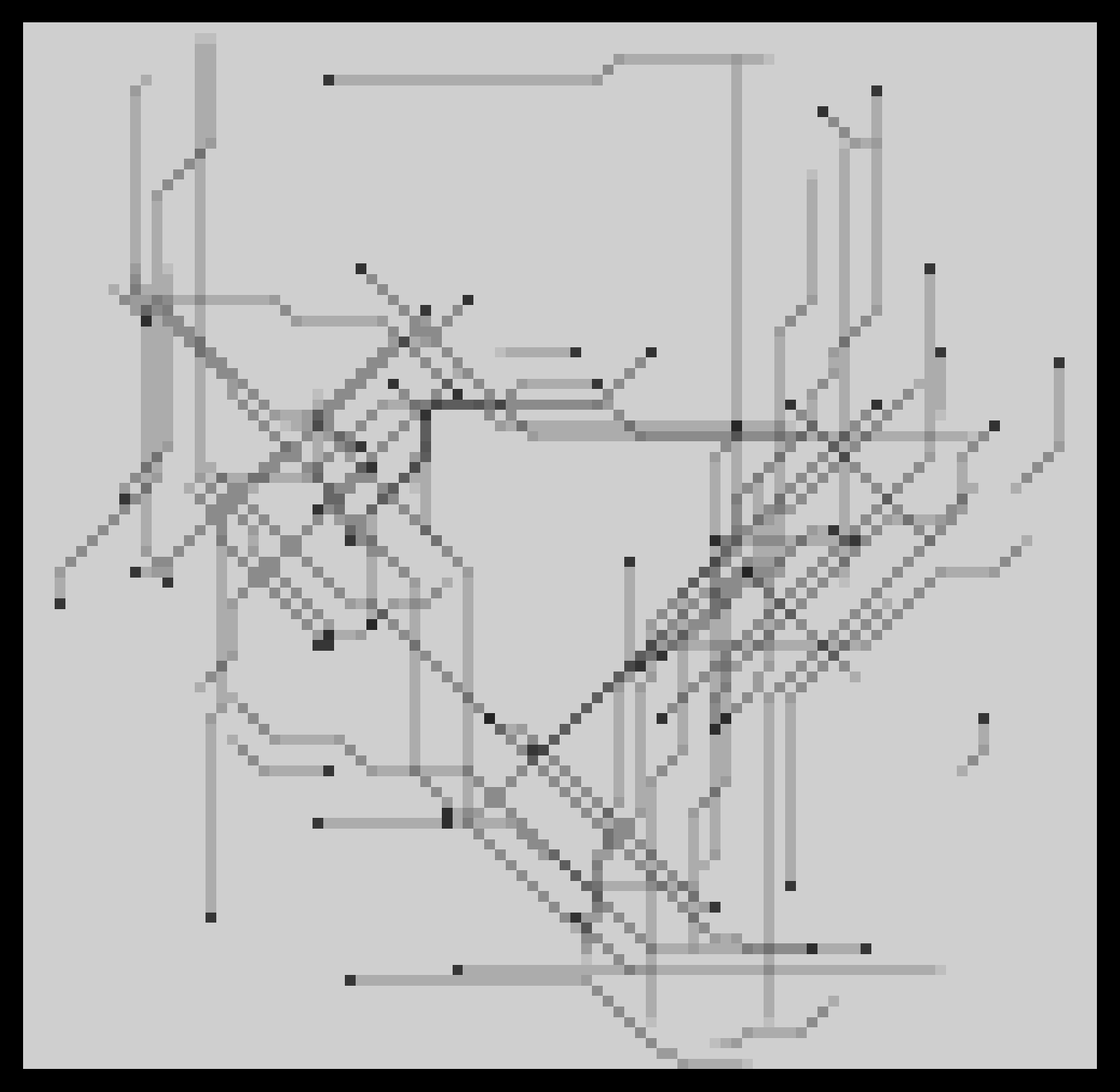}}
	\end{minipage}\\ \hline
    \end{tabular}
    \end{table}

\section{Conclusions}
\label{S:6}
    The problem of conflict-free path planning for UAM operation is studied in this paper. Literature investigation shows that very few methods consider efficient 4D usage of airspace from the perspective of traffic management, and existing methods lack a combination of the utilization of both time and space dimensions to achieve conflict resolution. In order to tackle this problem, we introduced and extended AirMatrix as a {concept} for four-dimensional airspace management. AirMatrix enables a unified framework for conflict avoidance in path planning. Based on this concept, shortest flight time path planning problem is formulated. And a CFA* algorithm is developed to solve the problem. 
    
    The CFA* algorithm includes a novel heuristic function that is suitable for AirMatrix concept, and a greedy algorithm employed for decision making while a dynamic obstacle is met during searching. The algorithm was realized and compared with original A* algorithm. The result shows that despite leading to flight delays, the CFA* algorithm resolved a large number of conflicts compared with A* algorithm, and has successfully provided a strategic layer of flight safety assurance. 
    
    The CFA* algorithm proposed in this study assumes an FCFS scheme, which reduces the complexity in 4D path planning, but potentially leads to inequality in airspace accessibility. An extension of this work might focus on a path planning algorithm for multiple flights with the same priority. Another potential extension of this study is a comprehensive safety analysis. This will count in uncertainties in UAM operation and lead to the quantitative sizing design of AirMatrix blocks.

\section*{Acknowledgement}
This research is supported by the Civil Aviation Authority of Singapore and the Nanyang Technological University, Singapore under their collaboration in the Air Traffic Management Research Institute. Any opinions, findings and conclusions or recommendations expressed in this material are those of the authors and do not reflect the views of the Civil Aviation Authority of Singapore. The authors also would like to thank Prof. Chen Lv for his insightful suggestions for this work. 

\bibliographystyle{elsarticle-num-names}
\bibliography{A Four Dimensional Urban Airspace Management Concept for Urban Air Mobility Operations.bbl}

\end{document}